\documentclass[apl, reprint, showpacs,citeautoscript,superscriptaddress,floatfix]{revtex4-2}
\usepackage{bm}
\usepackage[T1]{fontenc}
\usepackage{amsmath,amsfonts,amssymb}
\usepackage{graphicx}
\usepackage{amssymb}
\usepackage{epstopdf}
\usepackage{color}

\makeatletter

\renewenvironment{widetext@grid}{%
  \par\ignorespaces
  \setbox\widetext@top\vbox{%
   \vskip15\p@
   \hb@xt@\hsize{%
    \leaders\hrule\hfil
    \vrule\@height6\p@
   }%
   \vskip6\p@
  }%
  \setbox\widetext@bot\hb@xt@\hsize{%
    \vrule\@depth6\p@
    \leaders\hrule\hfil
  }%
  \onecolumngrid
  \let\set@footnotewidth\set@footnotewidth@ii
}{%
  \par
  \twocolumngrid\global\@ignoretrue
  \@endpetrue
}%

\makeatother

\usepackage[separate-uncertainty=true]{siunitx}
\usepackage{physics}
\usepackage{natbib}
\usepackage{ulem}
\usepackage[section]{placeins}
\usepackage{dblfloatfix}
\def \mupl {\MakeLowercase{\textmu}-PL }
\def \mum {\micro\meter}
\def \nm {\nano\meter}
\def \ev {\electronvolt}
\def \ns {\nano\second}
\def \wcm {\watt\per\square\centi\meter}
\def \gsl {\text{\textsl{g}}}

\graphicspath{{Figures/}}

\begin{document}

\title{Single photon emission and recombination dynamics  in  self-assembled GaN/AlN quantum dots}

\author{Johann Stachurski}
\email{johann.stachurski@epfl.ch}
\affiliation{Institute of Physics, \'Ecole Polytechnique F\'ed\'erale de Lausanne, EPFL, CH-1015 Lausanne, Switzerland}
\author{Sebastian Tamariz}
\affiliation{Institute of Physics, \'Ecole Polytechnique F\'ed\'erale de Lausanne, EPFL, CH-1015 Lausanne, Switzerland}
\affiliation{Current address: Université Côte d’Azur, CNRS, CRHEA, F-06560 Valbonne, France}
\author{Gordon Callsen}
\affiliation{Institute of Physics, \'Ecole Polytechnique F\'ed\'erale de Lausanne, EPFL, CH-1015 Lausanne, Switzerland}
\affiliation{Current address: Institut für Festkörperphysik, Universität Bremen, 28359 Bremen, Germany}
\author{Rapha\"el Butt\'e}
\affiliation{Institute of Physics, \'Ecole Polytechnique F\'ed\'erale de Lausanne, EPFL, CH-1015 Lausanne, Switzerland}
\author{Nicolas Grandjean}
\affiliation{Institute of Physics, \'Ecole Polytechnique F\'ed\'erale de Lausanne, EPFL, CH-1015 Lausanne, Switzerland}

\begin{abstract}

III-nitride quantum dots (QDs) are a promising system actively studied for their ability to maintain single photon emission up to room temperature. Here, we report on the evolution of the emission properties of  self-assembled GaN/AlN QDs for temperatures ranging from \SI{5} to \SI{300}{\kelvin}. We carefully track the photoluminescence of a single QD and measure an optimum single photon purity of $\gsl^{(2)}(0) = 0.05 \pm 0.02$ at \SI{5}{\kelvin} and $0.17 \pm 0.8$ at \SI{300}{\kelvin}. We complement this study with temperature-dependent time-resolved photoluminescence measurements (TRPL) performed on a QD ensemble to further investigate the exciton recombination dynamics of such polar zero-dimensional nanostructures. By comparing our results to past reports, we emphasize the complexity of recombination processes in this system. Instead of the more conventional mono-exponential decay typical of exciton recombination, TRPL transients display a bi-exponential feature with short- and long-lived components that persist in the low excitation regime. From the temperature insensitivity of the long-lived excitonic component, we first discard the interplay of dark-to-bright state refilling in the exciton recombination process. Besides, this temperature-invariance also highlights the absence of nonradiative exciton recombinations, a likely direct consequence of the strong carrier confinement observed in GaN/AlN QDs up to \SI{300}{\kelvin}. Overall, our results support the viability of these dots as a potential single-photon source for quantum applications at room temperature.

\end{abstract}

\maketitle

\section*{Introduction\label{Sec1:Intro}}

In the framework of the second quantum revolution \cite{Dowling2003}, single photon emitters (SPEs) have emerged as an important building block for the implementation of fast operating quantum devices. For such use, an ideal SPE should allow for the production of bright photon streams of high purity along with on-chip integration capabilities for industrial purposes. To date, there have been promising candidates in a wide range of solid-state systems \cite{Aharonovich2016}, e.g., silicon or nitrogen vacancies in diamond \cite{Kurtsiefer2000,Aharonovich2014,Bogdanov2018}, but also other localized defects in 2D \cite{He2015,Tran2016} and 3D \cite{Berhane2017,Zhou2018} semiconductor materials, as well as semiconductor quantum dots (QDs) \cite{Senellart2017,Holmes2019,Arakawa2020}.

While the development of III-nitride (III-N) QDs for SPE purposes is partially hindered by the dephasing induced by strong phonon \cite{Ostapenko2012} and point defect \cite{Bardoux2006,Gao2017} coupling in wurtzite QDs, recent research has revealed structures in which the impact of point defects is significantly reduced \cite{Arita2017}. Weaker dephasing can also be tailored either via a reduction in the QD size \cite{Honig2013,Kindel2014} or the absence of any built-in electric field in zinc-blende III-N QDs \cite{Fonoberov2003,Sergent2013}. In addition, III-N QDs remain of utmost relevance for room temperature (RT) SPE applications, where their state of the art III-arsenide counterparts are inoperable \cite{Wang2019}. III-N QDs may be particularly well-suited for quantum operations that do not require photon indistinguishability \cite{Aharonovich2016,Castelletto2020}, such as quantum-key distribution \cite{Scarani2009,Lo2014}, quantum imaging \cite{White1998} or quantum metrology \cite{Cheung2007}. Furthermore, the III-N system benefits from an already well-developed infrastructure resulting from its wide use for solid-state lighting applications, which translates into excellent epitaxial growth control and efficient bipolar doping \cite{Deshpande2013}. These examples of appealing properties render the III-N QD platform well adapted for broadband applications with QD emission ranging from the deep ultraviolet down to the infrared \cite{Damilano1999,Frost2013,Verma2014,Reilly2019}. In this regard, to date, RT single-photon emission based on III-N quantum heterostructures has already been demonstrated with both GaN/AlGaN \cite{Holmes2014,Holmes2016} and GaN/AlN QDs \cite{Tamariz2020}.

In the present work, we report on several SPE features up to RT of polar self-assembled (SA) GaN/AlN QDs grown on cost-effective Si(111) substrates that are potentially suitable for future on-chip integration. These QDs have been characterized using micro-photoluminescence ({\textmu}-PL) measurements under quasi-resonant excitation that were complemented by an analysis of their photon emission statistics. Second-order autocorrelation function ($\gsl^{(2)}(\tau)$) measurements have been carried out on several QDs as a function of excitation power density from 5 to \SI{300}{\kelvin} and analyzed in the framework of a multi-excitonic model. This allowed us to evaluate the impact of thermal broadening on the single photon purity as well as the limits of the adopted framework.

Furthermore, in order to clarify the origin of the large exciton decay times extracted from $\gsl^{(2)}(\tau)$ measurements, we performed complementary time-resolved photoluminescence (TRPL) experiments on an ensemble of QDs issued from the same sample over the same temperature range. We specifically investigated the low excitation regime and confronted the observed bi-exponential decays to the current understanding of exciton dynamics occurring in polar SA GaN/AlN QDs.

This article is structured as follows. In Sec.~"Main electronic features of polar III-nitride self-assembled quantum dots", we first recall some specific features of polar SA GaN/AlN QDs with a focus on their main electronic properties described in the framework of the hybrid biexciton model \cite{Honig2014} which is considered to explain the \mupl results obtained on such single QDs. Then, in Sec.~"Framework of the single photon emission measurements" we detail the multi-excitonic model that successfully accounts for the second order auto-correlation measurements performed on single QDs. In this regard, we also provide in Sec.~"Single photon emission from self-assembled quantum dots" a full description of the employed $\gsl^{(2)}(\tau)$ fitting function as well as an analysis of the experimental $\gsl^{(2)}(\tau)$ traces collected over two single QDs from \SI{5}{\kelvin} to RT. In Sec.~"Recombination Dynamics in a quantum dot ensemble", we give an overview of the time-dependent photoluminescence features of GaN/AlN QDs as opposed to conventional non-polar SPEs, before providing an in-depth analysis of the temperature- and power-dependent behavior of exciton recombination in Sec.~"Exciton recombination processes". To this end, we propose different scenarios to account for the observed multi-exponential decay and discuss their respective limits. In Sec.~"Review of excitonic decay in GaN/AlN SA QDs", we compare the exciton lifetimes extracted from TRPL transients to former results reported on SA GaN/AlN QDs of both experimental and theoretical nature. We finally conclude in Sec. "Discussion". Details about the QD growth procedure and the sample preparation for performing single dot spectroscopy are provided in Sec.~"Materials and methods", along with a brief mention of some relevant aspects regarding optical spectroscopy that are further detailed in the supplementary information (SI).

\section*{Results}

\subsection*{Main electronic features of polar III-nitride self-assembled quantum dots}\label{Subsec2.1:mainfeatures}

\begin{figure}
	\centering
	\includegraphics[width= 0.48\textwidth]{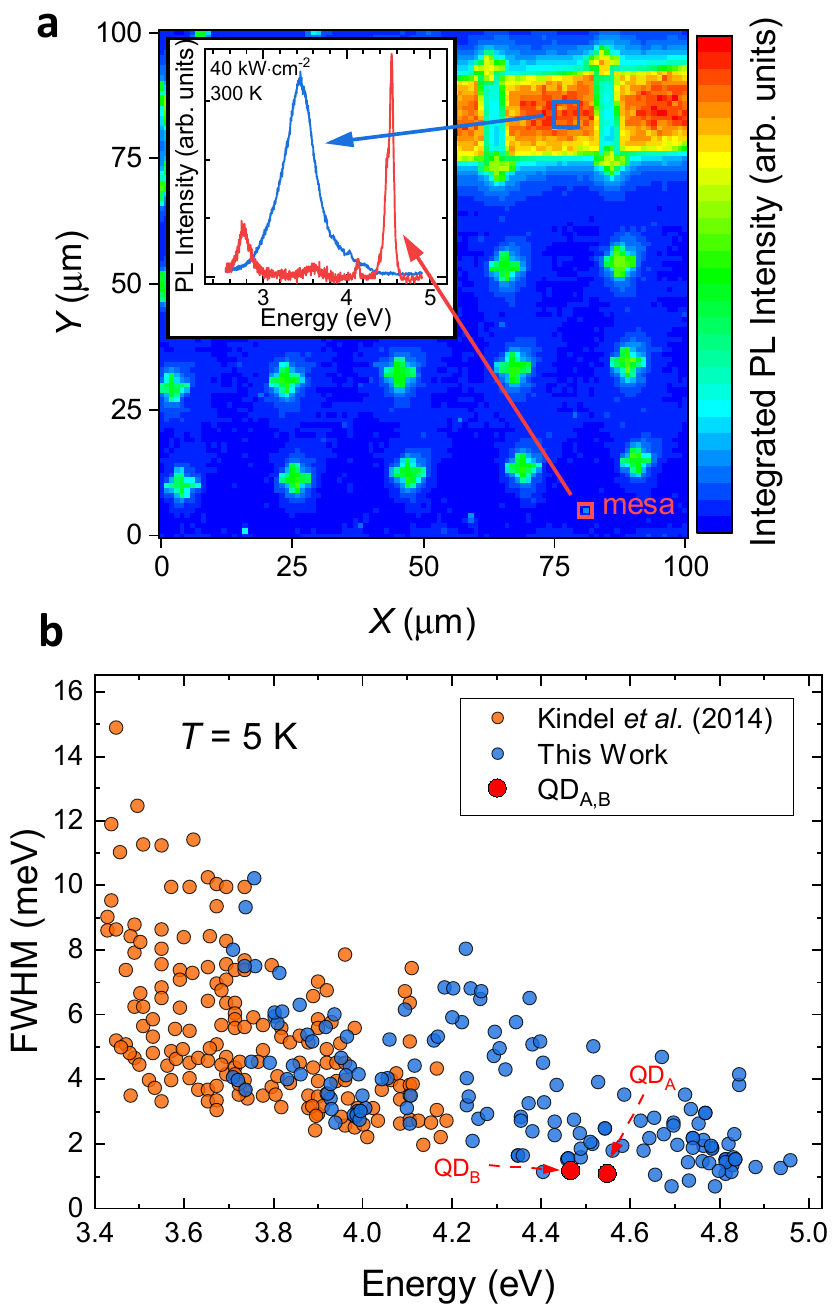}
	\caption{\textbf{a.} $100 \times 100 \ \si{\square\mum}$ micro-photoluminescence mapscan of the patterned sample for which the signal is recorded at \SI{300}{\kelvin} and integrated between 2.5 and \SI{4.5}{\ev}. Spatially-resolved micro-photoluminescence integrated intensity images are recorded using steps of one \SI{}{\mum}. The strongest signal is measured from unprocessed areas with a high QD density. Inset: Typical \mupl spectrum collected from an unprocessed area (blue line) and from a mesa (red line).
	\textbf{b.} Full width at half-maximum (FWHM) distribution of GaN/AlN excitonic emission recorded at low temperature ($T=$ 5 K) in the cw low excitation regime. Orange dots correspond to QDs grown by metalorganic vapor-phase epitaxy on SiC studied by Kindel \textit{et al.} \cite{Kindel2014}. All blue data points were extracted from the same sample. Auto-correlation measurements described in the following are performed on the QDs labelled QD$_{\rm A}$ and QD$_{\rm B}$}.
	\label{fig1:intro}
\end{figure}

 All the measurements reported in the present work were carried out on SA QDs grown by molecular beam epitaxy (MBE). The sample surface was patterned into mesa structures to ease the investigation of single QDs. Aspects related to the growth and subsequent processing of the sample are detailed in "Materials and methods", along with the quasi-resonant excitation scheme. The QD density varies between a few to about one hundred QDs per mesa. The vast majority of the dots emits around \SI{3.5}{\ev} at 300 K, as can be observed by recording \mupl spectra over unprocessed areas (Fig.~\ref{fig1:intro}\textbf{a}). These SA GaN/AlN QDs have a well-known truncated hexagonal pyramid shape inherited from the wurtzite crystal symmetry \cite{Arlery1999,Andreev2000,Butte2008}. Owing to the strong built-in electric field that arises along the polar \textit{c}-axis, trapped electrons get pushed toward the top of the pyramid while holes sit in the wetting layer (WL) part \cite{Andreev2000,Butte2008}. The strain induced by the GaN-AlN lattice mismatch results in a piezoelectric field component that leads to an additional lateral confinement of electron and hole wave functions \cite{Andreev2000,Ranjan2003}. 
The larger the QD height, the larger the Stark shift while the quantum confinement is weakened.  As a result, the exciton emission energy is redshifted. This quantum-confined Stark effect (QCSE) is experienced without any external electric field. Simultaneously, the electron-hole wave function overlap gets significantly reduced, hence leading to radiative lifetime variations by several orders of magnitude for a QD height increased by a few nanometers only \cite{Andreev2001,Schliwa2014}. The smallest QDs are thus the brightest, and as such, they constitute the main source of interest for the present work.

The strong dipole moment that stems from the electron-hole wave function separation also causes the emission energy of excitonic states to be quite sensitive to the electronic environment. Therefore, GaN/AlN QD PL lines are observed to broaden way above the natural linewidth of excitonic states. While this appears detrimental for quantum applications at cryogenic temperature, it can be used to estimate the density of defect states in the vicinity of QDs (see SI Sec. [S4-S5] for further details). Smaller QDs are less impacted by such perturbations, as illustrated by the decrease in the exciton linewidth with increasing emission energy (Fig.~\ref{fig1:intro}\textbf{b}). On the other hand, carriers confined in GaN/AlN QDs experience a large trapping potential, which stems from the very thin GaN WL thickness and the large band offset between AlN and GaN binary compounds. This ensures that high energy GaN/AlN QDs remain efficient optical emitters up to RT with a large energy separation between confined energy levels.

\begin{figure}
	\centering
	\includegraphics[width = 0.48\textwidth]{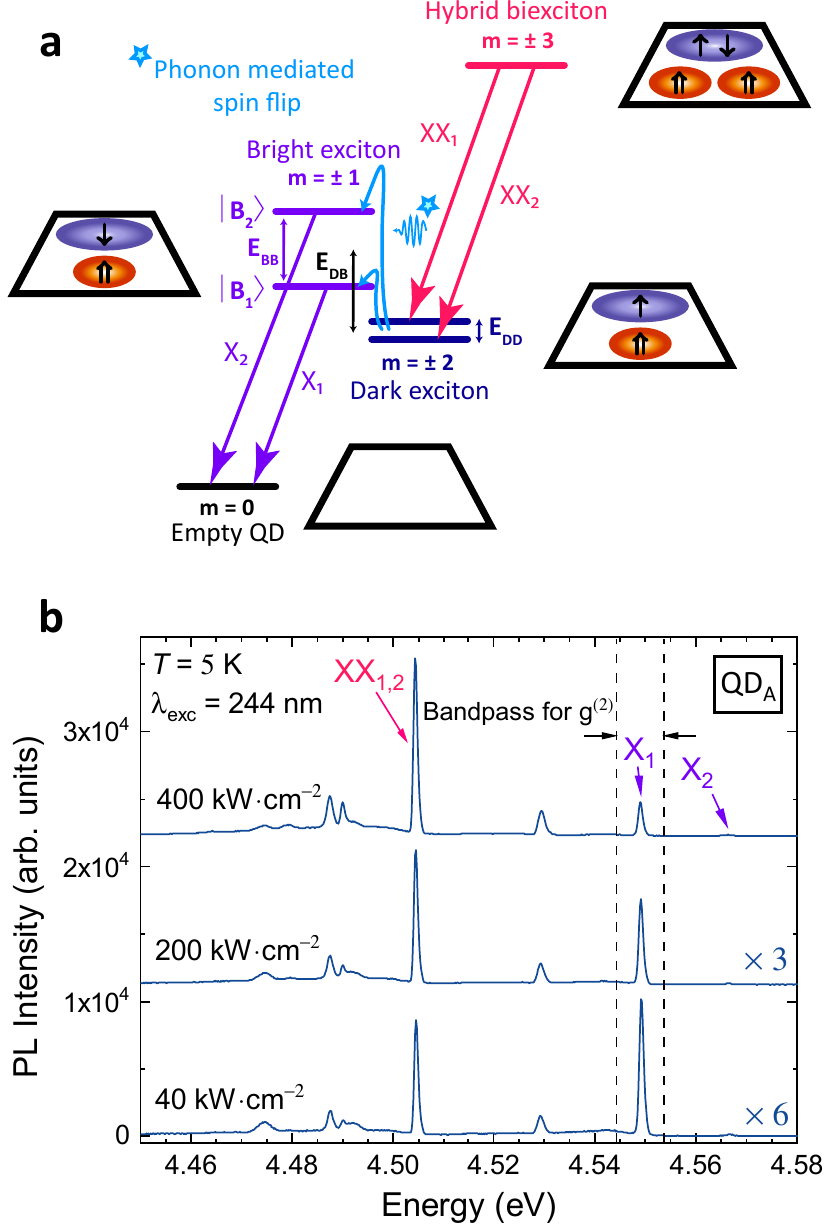}
	\caption{\textbf{a.} Schematic illustration of the hybrid exciton-biexciton cascade. \textbf{b.} Power-dependent \mupl spectra of QD$_{\rm A}$ exhibiting X$_{1,2}$ and XX$_{1,2}$ transitions recorded at 5 K.
	}
	\label{fig2:Pseries}
\end{figure}

The SA QDs we studied display a large base of a few tens of nanometers \cite{Widmann1998,Hoshino2004} and a height of 1 to \SI{3}{\nano\meter}. As a result of this low aspect ratio, the current picture of the electronic energy levels for these dots stands as follows: when two electron-hole pairs lie in the QD ground state, a pronounced in-plane spatial separation of the biexciton (XX) hole wave functions at the QD bottom prevails, which is enhanced by the large effective hole mass \cite{Rodina2001}. This wave function separation favors the XX configuration with parallel hole spins, which results in a huge binding energy that can reach tens of meV \cite{Honig2014}. This XX is referred to as a hybrid-XX and has a total angular momentum $m = \pm 3$.  The exciton ground state splits into two bright states ($m = \pm 1$, purple energy levels in Fig.~\ref{fig2:Pseries}\textbf{a}) and two dark states ($m = \pm 2$, dark blue energy levels in Fig.~\ref{fig2:Pseries}\textbf{a}) whose associated degeneracies are lifted likely due to symmetry reduction induced by QD in-plane elongation \cite{Ivchenko1977}. Fine structure splittings of a few meVs are usually observed in the polar SA GaN/AlN QD system \cite{Honig2014,Kindel2010} as a consequence of the strong built-in electric field \cite{Seguin2006}. The dark ($E_{\rm DD}$), bright ($E_{\rm BB}$) and dark-to-bright ($E_{\rm DB}$) state splittings are illustrated in Fig.~\ref{fig2:Pseries}\textbf{a}. $E_{\rm BB}$ is readily accessible in \mupl spectra as it is defined as the splitting between the two cross-polarized excitonic lines X$_{1,2}$ (purple arrows in Fig.~\ref{fig2:Pseries}\textbf{a}) shown in Fig.~\ref{fig2:Pseries}\textbf{b} for the QD labelled QD$_{\rm A}$. The linear polarization degree of X$_{1,2}$ exceeds \SI{90}{\percent} for all the investigated QDs (See SI Sect. [S3] for more details regarding single QD polarization measurements). The dark state splitting $E_{\rm DD}$ is much smaller and can be deduced from the splitting between the biexciton lines XX$_{1,2}$ (pink arrows). In most experimental situations, however, this splitting cannot be energetically resolved and a single biexciton line is observed \cite{Honig2014,Kindel2010}. The dark-to-bright state transition occurs through a phonon-mediated spin-flip process, whose contribution follows a Bose-Einstein distribution \cite{Honig2014,Liu2013}. Hence, at low temperature, only the low-energy bright state (B$_1$) gets significantly populated. For \mupl results, this translates into a bright X$_1$ line and a dim X$_2$ line. The recombination of the high-energy bright state (B$_2$), however, is characterized by a stronger oscillator strength \cite{Honig2014}, and the X$_2$ transition takes over with increasing temperature when the phonon bath involved in the dark-to-bright state transition gets populated (see Fig.~S5 in the SI for further details). Additional QD emission lines are commonly observed in \mupl spectra that are still under investigation. A more exhaustive description of the exciton-biexciton cascade can be found in the work by Hönig \textit{et al.} \cite{Honig2014}.

\subsection*{Framework of the single photon emission measurements \label{Subsec2.2:framework}}

The complexity of the multi-excitonic recombination scheme led us to methodically adapt the second-order autocorrelation $\gsl^{(2)}(\tau)$ function used to describe our results. Indeed, the customary two-level second order correlation function formula $\text{\textsl{g}}^{(2)}(\tau) = 1 - e^{-(\Pi+\gamma)\cdot\tau}$ can only be applied under low excitation conditions, for which the pump rate ($\Pi$) is smaller than the recombination rate ($\gamma$) \cite{Regelman2001}. Under higher excitation conditions, the fast relaxation process of excitons into the QD leads to a bunching phenomenon \cite{Mizrahi2003} that can be explained by the interplay of multi-carrier states: the QD is on average populated with several electron-hole pairs, so that the probability to witness the recombination of a single exciton is enhanced shortly after unloading the QD ground state energy levels. This bunching can be well accounted for by considering a simplified model of multi-excitonic processes \cite{Regelman2001}. In the latter model, transitions from and toward a level $\ket{n}$ of $n$ excitons trapped in the lowest possible QD energy states correspond to the recombination of a single electron-hole pair describing the QD relaxation toward level $\ket{n-1}$ with rate $\gamma_n$ and the capture of an additional electron-hole pair from level $\ket{n+1}$ with the pump rate $\Pi$ \cite{Dekel2000}, respectively. Assuming that linear scaling of the recombination rates holds, i.e., $\gamma_n = n\cdot\gamma_1=n\cdot\gamma$, the second order exciton correlation function can be expressed by a closed form expression \cite{Kindel2010a} that reads
\begin{equation}\label{multi-exc}
\gsl^{(2)}_{\rm X}(\tau) = \exp(\mu\cdot e^{-\gamma\abs{\tau}})\cdot\qty(1 - e^{-\gamma\abs{\tau}}),
\end{equation}
where $\mu = \Pi/\gamma$ is the QD mean occupation number.

In practice, excitonic states lying above the biexciton one are not observed in \mupl spectra upon increasing excitation power density for small QDs emitting above \SI{4}{\ev}. This could be explained by a saturation of the pump rate. This is modelled by considering the capture time of an exciton ($1/\Pi$) as the sum of the time needed to trap an exciton into an excited state, $\tau_{\rm t}$, and the relaxation time toward the exciton ground state ($\tau_{\rm r}$) \cite{Sun2020}. Variations in the excitation power density $P_{\rm exc}$ are only impacting the trapping time, such that $1/\tau_{\rm t} = \alpha \cdot P_{\rm exc}$, where $\alpha$ accounts for the pumping efficiency. The mean occupation number can thus be fitted as a function of excitation power density using a relationship given by:
\begin{equation}\label{sat}
\mu(P_{\rm exc}) = \frac{\mu_{\rm sat}}{1 + P_{\rm sat}/P_{\rm exc}},
\end{equation}
with $\mu_{\rm sat} = (\tau_{\rm r}\cdot \gamma)^{-1}$ and $P_{\rm sat} = (\alpha\cdot\tau_{\rm r})^{-1}$ being two fitting parameters, which leads to the expected linear dependence for low excitation conditions. 
Let us note here that Auger-assisted relaxation processes \cite{Ohnesorge1996,Ferreira1999} that could decrease $\tau_{\rm r}$ are not considered in this framework. Indeed, the latter phenomenon can be expected to be relatively weak considering our quasi-resonant excitation scheme for which interactions with excited carriers in the WL are unlikely. As a result nonradiative Auger recombination processes have not been reported in GaN/AlN QDs so far. However, we cannot fully exclude its contribution to account for the saturation of $\mu$ and the quenching of multi-excitonic states.

In addition to the impact of multi-excitonic states on photon statistics described above, spectral jittering of the exciton line occurring on the nanosecond to picosecond timescale could induce some additional photon bunching in the $\gsl^{(2)}(\tau)$ traces \cite{Gao2019,Holmes2021}. The amplitude of this bunching depends on the overlap between the emission line and the detection window and cancels out when the former is fully encompassed in the latter \cite{Sallen2010}. Thanks to the large energy spacing between the QD emission lines and the low background noise level, we could successfully avoid the impact of spectral wandering at low temperature by making use of an \SI{8}{\milli\ev} detection bandpass. The exciton line, whose linewidth amounts to \SI{1}{\milli\ev}, is thus fully encompassed in  this bandpass, as illustrated in Fig.~\ref{fig2:Pseries}\textbf{b}. At higher temperatures, the line broadening is mainly driven by phonon coupling occurring on a picosecond timescale, i.e., below the setup detection limit \cite{Fischer1997,Carmele2011,Callsen2013}. We can therefore expect spectral wandering of the exciton line to have no significant impact on the determination of the second order correlation function, regardless of the temperature.

\begin{figure}[h!]
	\centering
	\includegraphics[width = 0.48\textwidth]{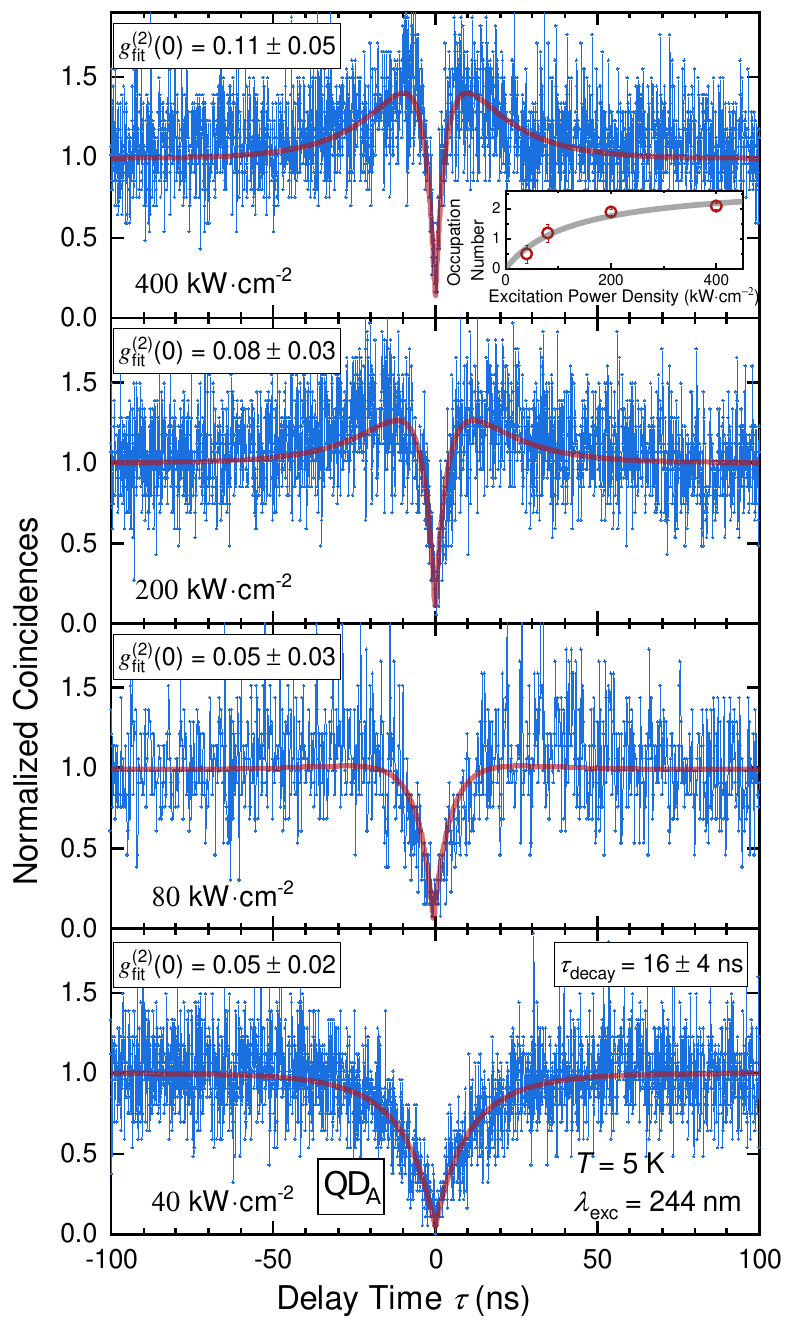}
	\caption{$\gsl^{(2)}(\tau)$ traces of QD$_{\rm A}$ recorded at 5 K as a function of excitation power density (connected blue diamonds) and convoluted fits (red lines). The channel resolution is fixed to 100 ps/channel at 40, 200 and \SI{400}{\kilo\wcm}. Given the smaller number of total coincidences measured at \SI{80}{\kilo\wcm}, the corresponding data are shown with a 200 ps/channel resolution. The exciton decay time being in theory power-independent, its value is extracted by fitting all traces with a shared $\tau_{\rm decay}$ value. The inset shows the evolution of the fitting parameter $\mu$ upon increasing power. The data (red circles) are fitted using Eq.\ \ref{sat} (grey line,) which yields $\mu_{\rm sat} = 2.9$. All uncertainties are obtained in the simplified framework described in the text and indicate the robustness of the fit. The full collected data span a range of $\pm$ \SI{130}{\ns} around 0.
	}
	\label{fig2bis:g2Pseries}
\end{figure}

Finally, we point out that the experimental fitting function also accounts for the instrument response function (IRF) such that
\begin{equation} \label{expg2}
\gsl^{(2)}_{\rm exp}(\tau) = \gsl^{(2)}_{\rm X}(\tau)\otimes \text{IRF}(\tau).
\end{equation}
The instrument response becomes detrimental only when the characteristic antibunching time approaches the time resolution of \SI{220}{\pico\second}.

\subsection*{Single photon emission from self-assembled quantum dots\label{Subsec2.3:g2results}}

$\gsl^{(2)}(\tau)$ traces recorded at various excitation power densities at $T=$ 5 K are displayed in Fig.~\ref{fig2bis:g2Pseries}. They all show a clear antibunching at zero delay time and a signature of the above-mentioned bunching phenomenon resulting from multi-exciton state occupancy under high excitation conditions. The purity determined with the fitting function ($\gsl^{(2)}_{\rm fit}$) is also given.
The  $\gsl^{(2)}(\tau)$ data are fitted using Eq.\ \ref{expg2} with a fixed exciton decay time $\tau_{\rm decay} = \gamma^{-1} = \SI{16(4)}{\nano\second}$ deduced by fitting all traces with a shared recombination rate $\gamma$. This value exceeds by about one to two orders of magnitude the decay times previously reported for GaN QDs emitting above the bulk GaN bandgap \cite{Daudin1999,Kako2003a,Bretagnon2006,Hrytsaienko2021}. First, this suggests a very low nonradiative recombination rate, as the QD of interest remains bright despite such a long-lived recombination time for the excitons. This comparatively long exciton decay time may originate from a transfer toward the dark states depicted in Fig.~\ref{fig2:Pseries}\textbf{a}, that could act as a long-lived reservoir in the absence of any efficient thermally enhanced phonon-mediated spin flip between dark and bright states \cite{Labeau2003,Sallen2009}. The single photon purity is observed to decrease with rising excitation power density, as a consequence of the temporal resolution. At \SI{40}{\kilo\watt\per\square\centi\meter}, the measured purity of \SI{0.05(2)}{} compares well to the lowest values measured to date in III-N systems, namely GaN QDs formed at step edges in low aluminum content GaN/AlGaN quantum wells \cite{Arita2017}. The narrowing of the antibunching dip can be explained based on an increase in the mean occupation number $\mu$, along with the pump rate. The sublinear pump power dependence of $\mu$ is shown in the inset of Fig.~\ref{fig2bis:g2Pseries} and highlights the saturation of the QD filling.
Let us note, however, that the fitting of the antibunching dip (red lines shown in Fig.~\ref{fig2bis:g2Pseries}) with a global variable $\gamma$ differs from the results obtained when fitting each $\gsl^{(2)}(\tau)$ trace individually. The decay time is especially affected by small variations in the mean occupation number, as a result of the exponential dependence in $\mu$ of the bunching (Eq.\ \ref{multi-exc}). This translates into a large uncertainty obtained for $\tau_{\rm decay}$ and a slight discrepancy between fitted and experimental traces. Variations in the measured exciton decay time could originate from power-dependent fluctuations of the built-in electric field. Such changes, ascribed to variations in the charging dynamics of defects surrounding the QDs \cite{Kindel2010a}, are, however, expected to be negligible for QD$_{\rm A}$ due to its high emission energy (see details in the SI). Alternatively, spin flip processes between dark and bright states could also account for changes in the exciton lifetime. A more realistic model would require to consider transitions rates between non-degenerate exciton and biexciton states that would add to the current number of unknown variables.

\begin{figure}
	\centering
	\includegraphics[width = 0.48\textwidth]{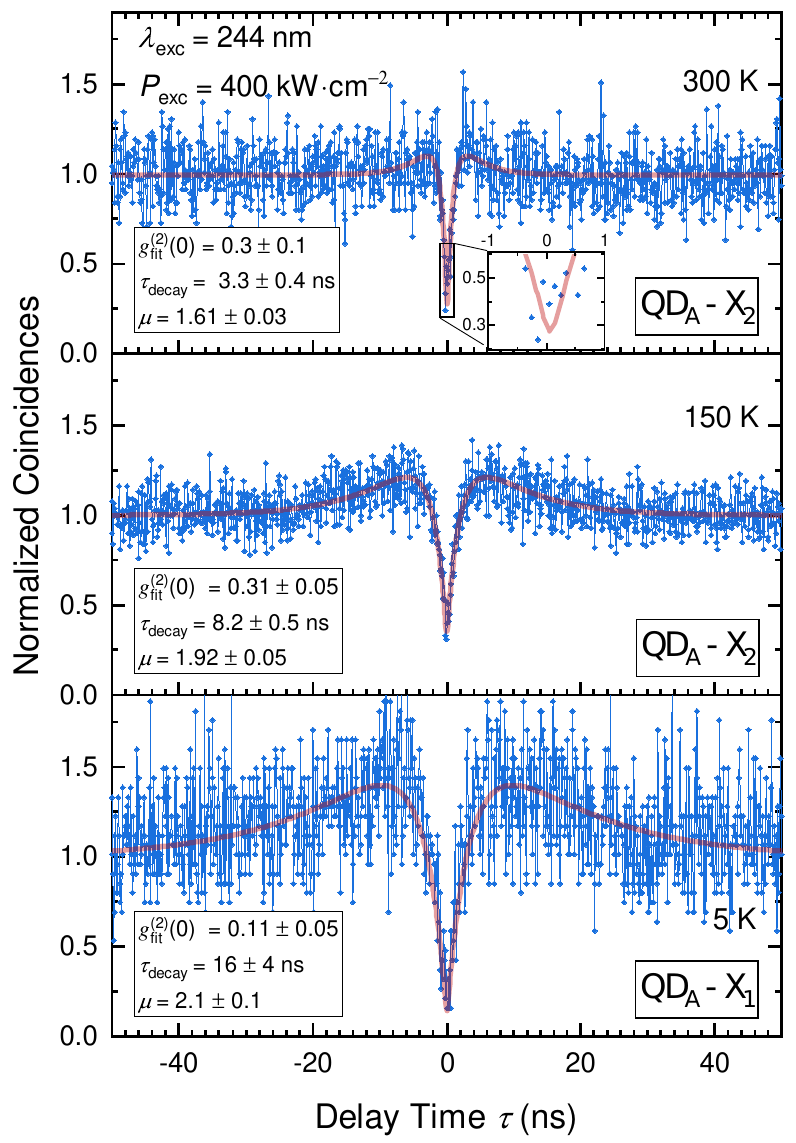}
	\caption{$\gsl^{(2)}(\tau)$ traces of QD$_{\rm A}$ recorded at an excitation power density of \SI{400}{\kilo\watt\per\square\centi\meter} as a function of temperature, with a channel resolution of 100 ps/channel (connected blue diamonds) and convoluted fits (red lines). All uncertainties are obtained in the simplified framework described in the text and only indicate the fitting robustness. The full collected data span a range  of $\pm$ \SI{130}{\ns} around 0. The inset in the top panel is a zoomed-in view of the antibunching dip.
	}
	\label{fig3:Tseries}
\end{figure}

In the following, we focus on the temperature-dependent behavior of SPE features exhibited by these SA GaN/AlN QDs. For such dots, it has been shown that the antibunching amplitude can display a progressive decrease above cryogenic temperature and cross the SPE limit ($\gsl^{(2)}(0)\geq$ 0.5) under the increased contribution of spurious background signals \cite{LeRoux2017}. However, the $\gsl^{(2)}(\tau)$ traces shown in Fig. \ref{fig3:Tseries}, measured successively on the X$_1$ (\SI{5}{\kelvin}) and X$_2$ lines (\SI{150}{\kelvin} and \SI{300}{\kelvin}), underline the remarkable conservation of the single photon purity from \SI{5}{\kelvin} up to RT for QD$_{\rm A}$. This can be explained by the large spectral separation between the exciton and the biexciton emission lines (see, e.g., Fig.~\ref{fig2:Pseries}\textbf{b}), such that the signal collected within the detector bandpass remains dominated by the exciton emission, despite the increasing weight of thermal broadening. The observed narrowing of the antibunching dip with increasing temperature is expected to result from both an increase in the exciton decay rate, as a consequence of both faster nonradiative recombination processes, and more presumably an increase in the pump rate as the phonon-assisted relaxation time $\tau_{\rm r}$ is reduced. In addition, in the specific case of these dots, the enhancement of the decay rate between \SI{5} and \SI{150}{\kelvin} is also explained by the larger oscillator strength of the X$_2$ transition that takes over the X$_{1}$ line with increasing temperature due to thermal population \cite{Honig2014}.

\begin{figure}
	\centering
	\includegraphics[width= 0.48\textwidth]{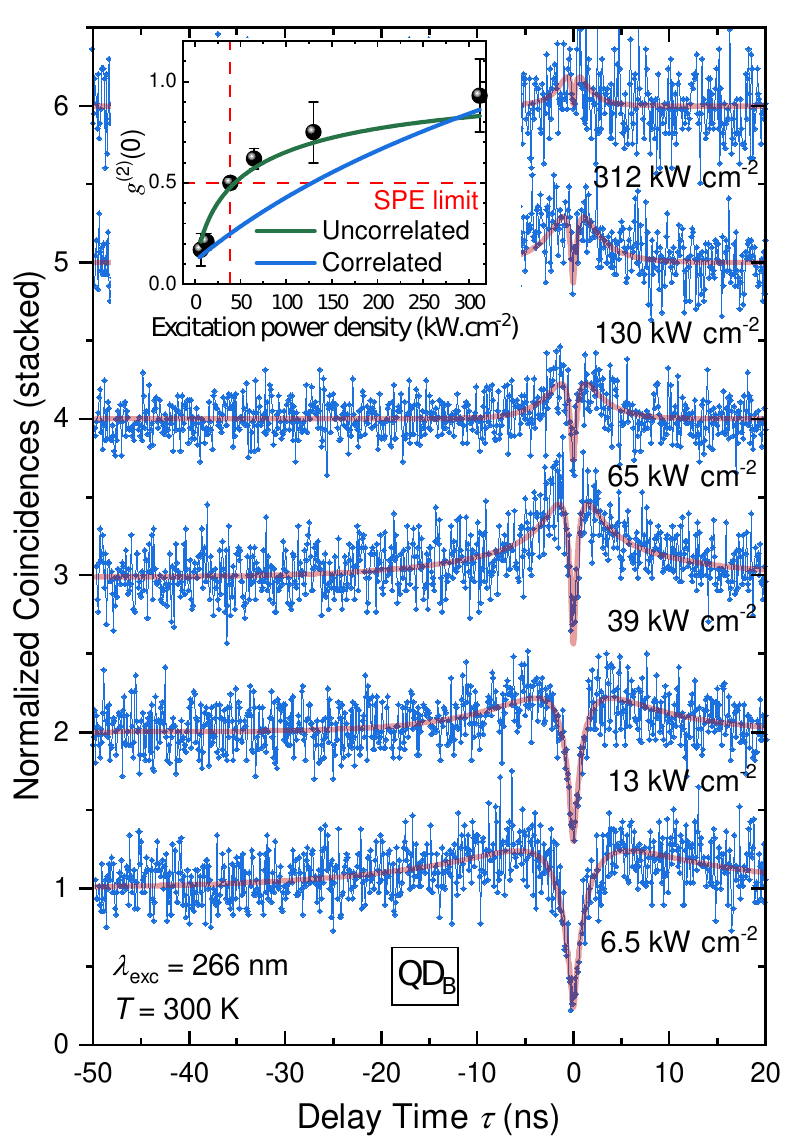}
	\caption{$\gsl^{(2)}(\tau)$ traces of QD$_{\rm B}$ recorded at \SI{300}{\kelvin} as a function of excitation power density with a channel resolution of 100\ ps/channel (blue connected diamonds) and convoluted fits (red lines). Inset: evolution of the single photon purity with rising excitation power density where the solid lines are the results issued from models assuming either an uncorrelated background (green line) or a correlated one (blue line). Adapted from Tamariz \textit{et al.} \cite{Tamariz2020}.	
	} 
	\label{fig4:S_paper}
\end{figure}

The high single photon purity observed for QD$_{\rm A}$ at RT is not an isolated case and has been measured for different QDs we investigated \cite{Tamariz2020}. To strengthen this point, Fig.~\ref{fig4:S_paper} displays the evolution of another SPE, labelled QD$_{\rm B}$, upon variation of the excitation power density, with a RT purity of $\gsl^{(2)}(0) = \SI{0.17(8)}{}$ at the lowest reported excitation power density of \SI{6.5}{\kilo\wcm}. The antibunching amplitude is again observed to decrease upon increasing excitation power density (see the inset of Fig.~\ref{fig4:S_paper}), which is in line with the superlinear increase in the biexciton intensity and its resulting stronger contribution to the recorded $\gsl^{(2)}(\tau)$ signal. Let us point out that the contrasting responses to excitation at equivalent power densities displayed by QD$_{\rm A}$ and QD$_{\rm B}$ could originate from strong changes in the absorption coefficient of such 0D nanostructures when excited at two different quasi-resonant excitation energies (4.66 and \SI{5.08}{\ev}).

\begin{figure}
	\centering
	\includegraphics[width= 0.48\textwidth]{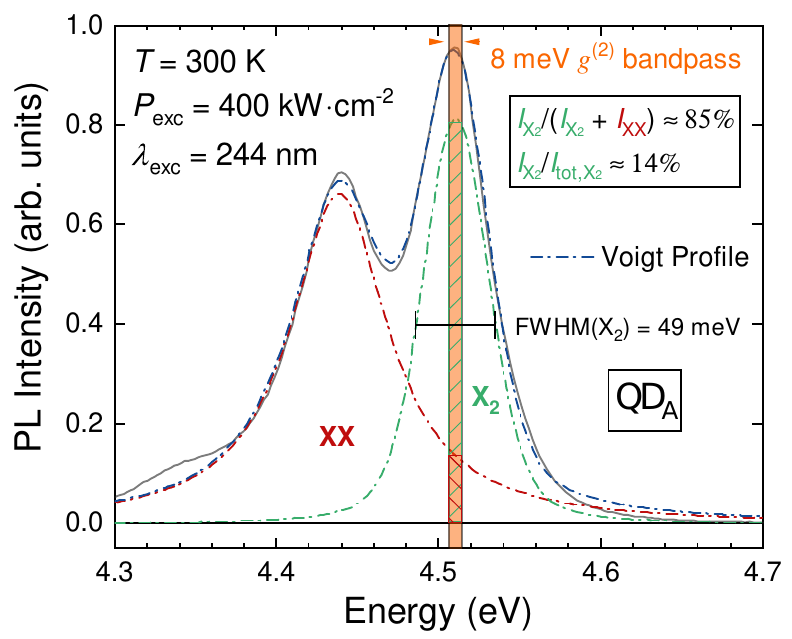}
	\caption{Room temperature \mupl spectrum recorded on QD$_{\rm A}$ approximated by a two-peak function. The biexciton is fitted with a Lorentzian (red dash-dotted line). X$_2$ is not fully Lorentzian and is best approximated by a Voigt function (green dash-dotted line), hinting at the underlying presence of X$_1$ whose impact remains however negligible. The resulting two-peak fit is highlighted with a blue dash-dotted line. The detection window of the HBT setup is indicated along with the contribution of X$_2$ ($I_{\rm X_2}$) and XX ($I_{\rm XX}$) to the measured signal (hatched areas). $I_{\rm tot,X_2}$ stands for the integrated intensity of the X$_2$ line. The FWHM of X$_2$ is highlighted.	
	} 
	\label{figA2:overlap}
\end{figure}

This loss in single photon purity is included in Eq.\ \ref{expg2} by introducing an offset parameter $\gsl^{(2)}(0)$ that accounts for the contribution of parasitic background emission. As previously mentioned, at RT, the strongest contribution to this background emission stems from the broadened XX emission line, which is overlapping with the detection window. The inset of Fig.~\ref{fig4:S_paper} shows the expected evolution of the single photon purity as a function of excitation power density, where the blue line is a theoretical estimate of $\gsl^{(2)}(0)$ when summing up all second order auto- and cross-correlation terms involving X and XX \cite{Tamariz2020}. The observed loss in single photon purity is eventually more properly described by considering the XX luminescence as an uncorrelated background (green line in the inset of Fig.~\ref{fig4:S_paper}) along the lines given in Ref. \cite{Brouri2000}:
\begin{equation}\label{uncorr}
\gsl^{(2)}(0) = 1 - \rho^2 \ \ ; \ \ \rho = \frac{I_X}{I_X + I_{XX}}.
\end{equation} $I_X$ and $I_{XX}$ correspond to the intensity of the X and the XX lines within the bandpass of the Hanbury-Brown and Twiss (HBT) interferometer setup (see "Materials and methods"), that are directly deduced from the related \mupl spectra as depicted in Fig.~\ref{figA2:overlap}. At \SI{400}{\kilo\wcm}, the signal-to-noise ratio $\rho \simeq 0.85$ measured on QD$_{\rm A}$ would lead to a single photon purity of 0.27, a value in very good agreement with experimental observations (Fig.~\ref{fig3:Tseries}). In this respect, a large exciton-biexciton splitting appears crucial to ensure single-photon emission at RT. The low $\gsl^{(2)}(0)$ value we report is explained by the X$_2$-XX splitting we measure that ranges from \SI{62}{\milli\ev} (\SI{5}{\kelvin}) to \SI{72}{\milli \ev} (\SI{300}{\kelvin}) for QD$_{\rm A}$. Let us note that this exciton-biexciton splitting is larger than the biexciton binding energies previously reported for GaN/AlN QDs \cite{Holmes2019}, which most likely stems from a particularly small QD aspect ratio ($\sim 0.1$) known to result in a large positive biexciton binding energy ($E_{\rm X}-E_{\rm XX} > 0$) \cite{Honig2014}. Qualitatively, a large aspect ratio (large QD diameter) leads to a reduced pairwise Coulomb repulsion, while the electron-hole exchange interaction is enhanced for small QDs \cite{Tomic2009}. On this basis, the observed energy variations in the biexciton binding energy with temperature could tentatively be ascribed to a complex interplay between the respective weight of direct Coulomb interaction and exchange, as well as correlation interactions experienced by trapped carriers upon lattice expansion. While the $\gsl^{(2)}(\tau)$ spectral bandpass of our HBT setup is limited to a maximum value of \SI{8}{\milli\ev}, the collection efficiency of the exciton photoluminescence could be improved by making use of a larger detection window while causing little deterioration to the single photon purity. As an illustration, $\rho$ would be reduced from $85\%$ to $82\%$ for QD$_{\rm A}$ at an excitation power density of \SI{400}{\kilo\watt\per\square\centi\meter} (Fig.~\ref{figA2:overlap}) for an increase in the $\gsl^{(2)}(\tau)$ bandpass from $\Delta E = \SI{8}{\milli\ev}$ to $\Delta E = \SI{49}{\milli\ev}$, the latter value corresponding to the exciton linewidth reported at RT. Hence, the corresponding 5\% loss in single photon purity would result in a 5 times brighter signal. More details are provided in the SI Sec.~[S6].

The large spectral separation between the exciton and the biexciton lines and the high exciton binding energy of GaN/AlN QDs make them particularly suited for RT single-photon emission, as highlighted by the single photon purity $\gsl^{(2)}(0) = \SI{0.17(8)}{}$. In this regard, the role of the very thin WL, with a high electronic state energy of about \SI{5.3}{\ev} (i.e., \SI{234}{\nm}),  is also crucial. Indeed, the \SI{244}{\nano\meter} and \SI{266}{\nano\meter} continuous-wave lasers used to pump the structure generate electron-hole pairs directly into the QDs, hence avoiding any spurious signal from the WL.

\subsection*{Recombination Dynamics in a quantum dot ensemble\label{Sec2.4:TRPLintro}}

\begin{widetext}

\begin{figure}[h]
	\centering
	\includegraphics[width= 1\textwidth]{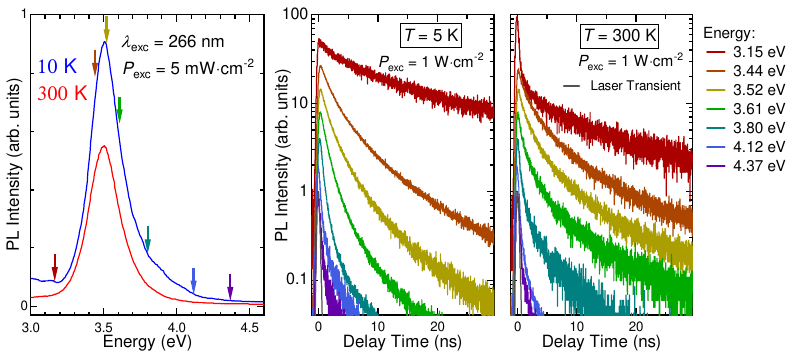}
	\caption{ \textbf{a.} Temperature dependence of the QD ensemble photoluminescence. The arrows mark the energies at which the TRPL transients shown in \textbf{b.} and \textbf{c.} are measured. The excitation power density is chosen low enough to ensure that most of the signal stems from excitonic recombination.  Time-resolved photoluminescence transients measured at \SI{5}{\kelvin} \textbf{b.} and \SI{300}{\kelvin} \textbf{c.} are displayed. The laser transient is also given (grey line).
	}\label{figTRPL1:Energy}
\end{figure}

\end{widetext}

While robust RT single-photon emission has been demonstrated for these SA GaN/AlN QDs, the long exciton decay time of \SI{16(4)}{\ns} extracted from second-order correlation function measurements recorded at \SI{5}{\kelvin} on QD$_{\rm A}$, whereas it is emitting near 4.5 eV, is not in line with previous experimental results \cite{Daudin1999,Kako2003a,Bretagnon2006,Hrytsaienko2021}. Therefore, in order to get a more general and more conclusive picture on the exciton dynamics in this system, we performed TRPL measurements on an ensemble of SA QDs by probing an unprocessed area on the sample. PL transients were collected for temperatures ranging from 5 to \SI{300}{\kelvin} and excitation power densities covering the 2 to \SI{1.2e3}{\milli\watt\per\square\centi\meter} range, leading to a maximum energy density per pulse of \SI{0.15}{\milli\joule\per\square\centi\meter}. While in the low density regime the PL spectra of the QD ensemble peak at $\sim$\SI{3.5}{\ev} whatever the temperature (see Fig.~\ref{figTRPL1:Energy}\textbf{a}), the PL intensity remains large enough to record PL transients of QDs emitting from 3.2 to \SI{4.5}{\ev}. 

PL transients measured at 5 and \SI{300}{\kelvin} are shown in Figs.\ \ref{figTRPL1:Energy}\textbf{b} and \ref{figTRPL1:Energy}\textbf{c} for various QD emission energies corresponding to a spectral window that ranges from \SI{4}{\milli\ev} for low energy emitting QDs to \SI{8}{\milli\ev} for high energy ones. At high excitation power density, all transients display a multi-exponential decay profile, which can be intuitively attributed to multi-carrier filling of the QDs. A contribution from the WL can, once again, be discarded as all the dots are pumped quasi-resonantly. Unlike previous observations \cite{Hrytsaienko2021}, the decay rate at short delays is strongly impacted by the QD size. In fact, the initial decay rate is enhanced for high energy emitters, hinting at an increase in the multi-carrier recombination rate due to the larger electron-hole wave function overlap. As a comparison, in colloidal QDs multi-carrier recombination is usually dominated by nonradiative Auger recombination processes occurring on a ps timescale \cite{Melnychuk2021}. On the other hand, Auger processes are commonly considered as negligible in self-assembled InAs/GaAs QDs and were only recently observed to occur on a timescale on the order of a few microseconds \cite{Kurzmann2016}. As such, multi-carrier recombination rates can be described assuming a pure radiative recombination process \cite{Santori2002}. We initially attempted to account for TRPL data in this framework, assuming the same linear scaling of the recombination rates ($\gamma_n = n\cdot\gamma_1=n\cdot\gamma$) than that considered to depict SPE results. However, when applying this approach to a QD ensemble the increasingly fast decay observed at high excitation power density could not be satisfactorily reproduced. This discrepancy could be attributed either to Auger processes speeding up the recombination of multi-carrier states or even faster multi-excitonic radiative recombination processes. One can also note that at high excitation and high emission energy the fast decays occur on a timescale on the order of a nanosecond or even below which is close to the laser pulse duration (Figs. \ref{figTRPL1:Energy}\textbf{b} and \ref{figTRPL1:Energy}\textbf{c}), hence preventing us from reaching any firm conclusion at this stage.

In addition, we recall that polar III-N QDs differ in a specific manner from their InAs/GaAs counterparts.
Indeed, QDs emitting below the bulk GaN bandgap experience a large QCSE leading to a power-dependent Stark shift \cite{Bretagnon2006}. This shift can in turn drive different emitters in and out of the detection window. This behavior stems from a progressive descreening of the built-in field when the number of trapped carriers decreases upon increasing time and could in turn additionally explain the failure of the above-mentioned modeling. The impact of the excitation power density on PL spectra is clearly discernible through a spectral shift of the QD ensemble peak energy by up to \SI{50}{\milli\ev} between 2 and \SI{1.2e3}{\milli\wcm} (see Fig.~S9 in the SI); a shift that vanishes for high energy QDs, which are less affected by the QCSE. Alternatively, QDs emitting above \SI{4}{\ev} exhibit large positive biexciton binding energies (${E_{\rm{XX}} < E_{\rm{X}}}$) that can reach a few tens of meV \cite{Honig2014}. Hence, the intermixing of different excitonic complexes could potentially occur when collecting TRPL transients at high energy that might affect in some ways the readability of the results.

\begin{figure}
	\centering
	\includegraphics[width= 0.48\textwidth]{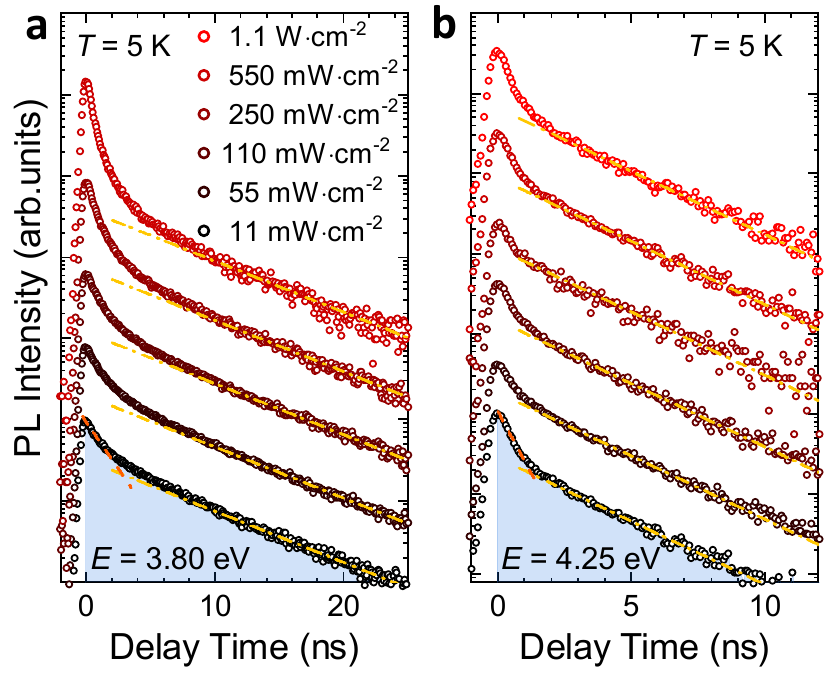}
		\caption{Photoluminescence decays of GaN QDs emitting at \textbf{a.} \SI{3.80}{\ev} and \textbf{b.} \SI{4.25}{\ev}, recorded at \SI{5}{\kelvin} for increasing excitation power densities. The tail of each transient is approximated by a single exponential with a $\tau_{\rm decay}$ value of 6.5 and \SI{2.8}{\ns}, respectively  (yellow dash-dotted lines). At low excitation power density, the decay is bi-exponential. The fast decay component is also highlighted in the graph (orange dashed line). The channel resolution was reduced down to 80 ps/channel to improve the qualitative appraisal of decay time conservation. As an illustration, the filled areas correspond to the integrated intensity $I_{\rm tot}$ collected at \SI{11}{\milli\wcm}.}
	\label{figTRPL2:Power}
\end{figure}

\subsection*{Exciton recombination processes\label{Subsec2.5:excitonTRPL}}

Given the above-mentioned complexity of multi-carrier processes, we can first focus on the description of the physics related to the tail of the TRPL transients instead. A long-lived mono-exponential decay is observed at any selected energy, which we associate to single exciton recombinations across the QD ensemble in the following. Figure~\ref{figTRPL2:Power} shows the power-dependent evolution of TRPL transients for QDs emitting at 3.80 and \SI{4.25}{\ev} recorded at a temperature of \SI{5}{\kelvin}. At low excitation power density, the long-lived mono-exponential decay is maintained over a dynamic range exceeding two orders of magnitude. The characteristic decay time ($\tau_{\rm decay}$) is equal to \SI{6.9(3)}{\ns} and \SI{2.8(2)}{\ns}, respectively, (cf. yellow dash-dotted lines in Fig.~\ref{figTRPL2:Power}). This long-lived decay is preserved at higher excitation power density and the tail of each transient can be consistently approximated by the same lifetime, as expected for the radiative recombination of excitons emitting at a given energy.

\begin{figure}
	\centering
	\includegraphics[width= 0.48\textwidth]{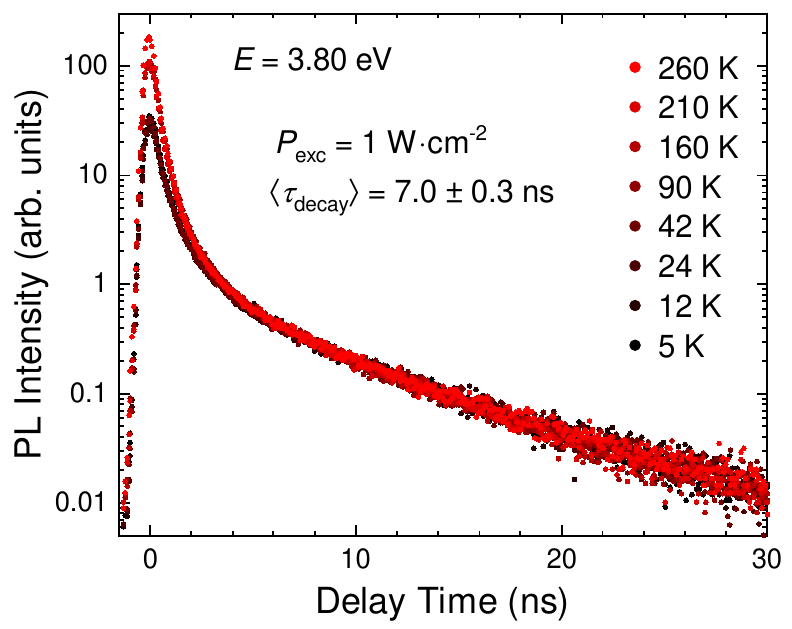}
	\caption{Photoluminescence decays of GaN QDs emitting at \SI{3.80}{\ev} recorded from \SI{5} to 260 K at an excitation power density of \SI{1}{\wcm}. All curves are stacked in order to superimpose the tail of the TRPL transients. The channel resolution was reduced down to 80\ ps/channel to improve the qualitative appraisal of decay time conservation. 
	}
	\label{figTRPL4:TemperatureDecays}
\end{figure}

To further support this picture associated to single exciton decay, low temperature TRPL measurements have been completed by a temperature dependent series for QDs emitting at the same energy of \SI{3.80}{\ev} (Fig.~\ref{figTRPL4:TemperatureDecays}). Interestingly, we do not observe any noticeable change in the decay time extracted from the tail of TRPL transients when heating the sample up to \SI{260}{\kelvin}. Such a behavior is expected for zero-dimensional nanostructures for which the radiative lifetime is independent of temperature \cite{Lasher1964}. We note that the same observation has been reported on SA GaN/AlN QDs grown on \textit{c}-plane sapphire substrates \cite{Renard2009decay}. This is also consistent with a system free from any nonradiative recombinations.

\begin{figure}
	\centering
	\includegraphics[width= 0.48\textwidth]{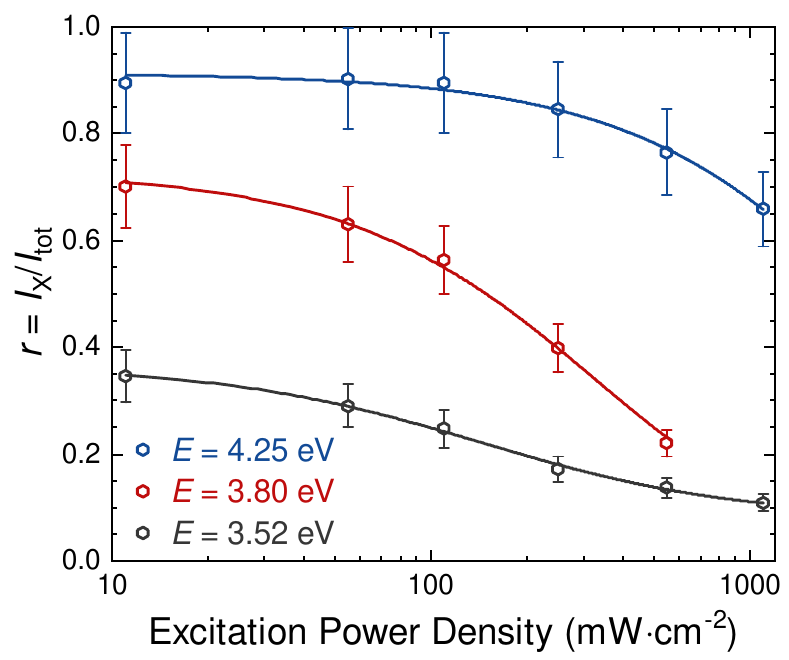}
	\caption{Low temperature ($T=$ 5 K) evolution of the $I_{\rm X}/I_{\rm tot}$ ratio as a function of excitation power density for various QD emission energies. Solid lines serve as a guide to the eye.
	}
	\label{figTRPL2bis:Ratios}
\end{figure}

With solid evidence that the tail of the TRPL transients is associated to single exciton decay, we can now compare the weight of this long-lived mono-exponential PL component to that of the whole decay profile at different energies and for varying excitation power densities. The exciton intensity $I_{\rm X}$ is estimated by integrating the mono-exponential fit for times following the disappearance of the short lifetime component of each transient. The results are normalized to the integrated intensity $I_{\rm tot}$ measured over the whole raw data ($\tau>0$) for each decay profile. The power-dependent trend of $r = I_{\rm X}/I_{\rm tot}$ for PL decay traces recorded at $T=$ 5 K is given in Fig.~\ref{figTRPL2bis:Ratios} for QDs emitting at 3.52, 3.80 and \SI{4.25}{\ev}. The weight of single exciton PL is logically observed to decrease with increasing excitation power density as multi-excitonic recombination gets more likely. Inversely, we would expect $r$ to converge to 1 as the average number of electron-hole pairs per QD drops below unity ($I_{\rm tot}(P_{\rm exc}\rightarrow 0) \rightarrow I_{\rm X}$). The latter hypothesis is experimentally invalidated as we witness a decrease in $r_0 = r(P_{\rm exc}=0)$ with decreasing QD emission energy. This manifests itself as a persisting bi-exponential decay, which does not vanish at low excitation power density (see Fig.~\ref{figTRPL2:Power}). However, the asymptotic saturation of $r$ toward $r_0$ upon decreasing excitation power density, which is most noticeable for high energy emitting QDs, is consistent with a picture where the observed PL decays at low power densities do not originate from multi-carrier recombination.

In this regard, transients collected at the lowest $P_{\rm exc}$ values can be described using the two-component decomposition:
\begin{equation}\label{biexp}
    I(t) = A_{\rm S}\cdot e^{-t/\tau_{\rm S}} + A_{\rm L}\cdot e^{-t/\tau_{\rm L}},
\end{equation}
with a short ($\tau_{\rm S}$) and a long ($\tau_{\rm L}$) decay term. $A_{\rm S,L}$ prefactors are fitting parameters, which can be linked to the ratios $r_0$ shown in Fig.~\ref{figTRPL2bis:Ratios} as follows:
\begin{equation}\label{biexpratio}
    r_0 = \frac{\tau_{\rm L}\cdot A_{\rm L}}{\tau_{\rm S}\cdot A_{\rm S} + \tau_{\rm L}\cdot A_{\rm L}} = \frac{\tau_{\rm L}}{\tau_{\rm S}\cdot A_{\rm S}/A_{\rm L} + \tau_{\rm L}}.
\end{equation}

At this stage, one should provide a physical explanation able to connect the short-lived component of PL decay traces to the long-lived one. On the basis of the schematic electronic picture shown in Fig. \ref{fig2:Pseries}\textbf{a}, two approaches could potentially account for the observed bi-exponential transient. As a first possibility, the bi-exponential decay could be attributed to interactions taking place between dark and bright states. In this framework, excitons trapped in a bright state can either recombine radiatively, with rate $\tau_S^{-1}$, or undergo a spin-flip process toward a dark state. The long-lived decay would thus relate to the reloading from the dark state with characteristic time $\tau_{\rm L}$ \cite{Sallen2009}. Yet dark-to-bright state transitions are phonon-mediated spin-flip processes. As such, they get enhanced when the number of phonons with high enough energy is increased, i.e., when the temperature increases. In this picture, we expect the reloading time $\tau_{\rm L}$ to decrease with temperature, down to a point where the exciton decay becomes mono-exponential, when $\tau_{\rm L}\ll \tau_{\rm S}$. This however does not agree with experimental observations, as we do not observe any change whatsoever in $\tau_{\rm L}$ when heating the sample (see Fig.~\ref{figTRPL4:TemperatureDecays}).

Alternatively, we could attribute the bi-exponential decay to exciton radiative recombination from the low energy ($\gamma_{\rm B_1}$) and high energy ($\gamma_{\rm B_2}$) bright states, respectively. At low temperature, most QD photoluminescence is supposed to originate from B$_{\rm 1}$, as reloading from dark states toward B$_{\rm 2}$ is thermally suppressed. However, radiative recombination from B$_{\rm 2}$ cannot be completely discarded and the fast decay could be interpreted as $\tau_{\rm S}^{-1} \simeq \gamma_{\rm B_2} + \gamma_{\rm 2\downarrow}$, where $\gamma_{\rm 2\downarrow}$ would correspond to the relaxation rate from B$_{\rm 2}$ to the dark states.

In this framework, we can foresee an enhancement in the B$_{\rm 2}$ luminescence either through a reduction in the bright state splitting $E_{\rm BB}$ or through a thermal enhancement in B$_{\rm 2}$ reloading. Experimentally, this can be linked to a decrease in $r_0$, i.e., an increase in the $A_{\rm S}/A_{\rm L}$ ratio. As shown in Fig.~\ref{figTRPL2bis:Ratios}, $r_0$ drops from 0.9 at \SI{4.25}{\ev} to 0.37 at \SI{3.52}{\ev}. Such an observation is consistent with the reported reduction in the bright-state splitting $E_{\rm BB}$ with decreasing QD emission energy \cite{Kindel2010}. In the same vein, we observe a drop from $r_0=0.7$ at \SI{5}{\kelvin} to $r_0=0.5$ at RT for QDs emitting at \SI{3.80}{\ev} (see Fig.~S10 in the SI), which is consistent with a thermal activation of the short-lived decay signal. However, due to the reduced intensity of the high energy tail of the QD ensemble PL, deriving the temperature-dependence of $r$ at various excitation power densities for QDs emitting above \SI{4}{\ev} proved a too challenging task. Nonetheless, we could still observe a drastic drop in the weight of the long-lived emission component of the transients for QDs emitting at \SI{4.25}{\ev} with increasing temperature, as $r$ decreased from 0.82 at {\SI{5}{\kelvin}} down to 0.25 at RT at an excitation power density of \SI{0.35}{\wcm}. Further details are provided in the SI Sec.~[S8].

Although the bi-exponential decay seems to be fairly well accounted for by exciton recombination from the two bright states, several considerations still challenge this hypothesis. First, the bright state splitting is expected to become negligible below \SI{4}{\ev} \cite{Kindel2010}, with a value dropping below \SI{1}{\milli\ev}. Hence, the refilling of B$_2$ should be fully activated for QDs detected at such energies. Conversely, the strong reduction of $r_0$ we measured from 3.8 to \SI{3.5}{\ev} suggests a weak contribution of B$_2$ recombination at \SI{3.8}{\ev}. Secondly, the lift of degeneracy between B$_1$ and B$_2$ must be accompanied by a convergence of their respective radiative recombination times. In practice, however, $\tau_{\rm L}$ was consistently measured to be 4 to 5 times larger than $\tau_{\rm S}$, regardless of the emission energy. From this last observation it clearly appears that the bi-exponential decay can no longer be solely explained by the recombination of the exciton bright states for low energy QDs. With this in mind, the B$_2$ phonon-mediated refilling mechanism described beforehand could also be associated to alternative QD states involving B-hole states. Indeed, so far, we only considered excitonic states for which the hole is located in the A- valence band. A more complete analysis could take into account states originating from both the A- and B- valence bands. The splitting between A- and B- exciton states has been theoretically estimated to amount to more than \SI{10}{\milli\ev} in GaN/AlN QDs, with little dependence on the QD size \cite{Winkelnkemper2008}. Hence, this splitting is large enough to strongly inhibit B-exciton emission at \SI{5}{\kelvin} while allowing its thermal activation at RT. In this perspective, the fast decay time $\tau_{\rm S}$ can be interpreted as the result of the intermixing of the B-exciton radiative recombination time and its relaxation time toward the A-exciton, even though the latter interpretation still lacks any direct experimental evidence.

At this stage, the analysis we propose here remains only partially satisfactory as it does not allow us to conclude unequivocally on the origin of the short-lived decay. Besides, reconciling the long decay time determined by autocorrelation measurements at \SI{5}{\kelvin} ($\tau_{\rm decay} = \SI{16(4)}{\ns}$) and the long-lived decays ($\tau_{\rm L}(\SI{4.5}{\ev}) \simeq \SI{3}{\ns}$) extracted from TRPL data remains a daunting challenge, as detailed hereafter.

\begin{figure}
	\centering
	\includegraphics[width= 0.48\textwidth]{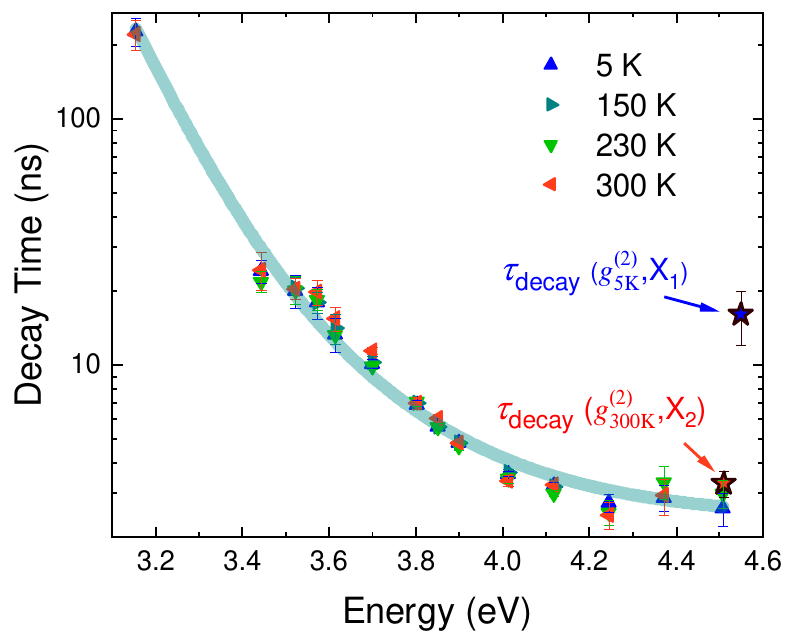}
	\caption{Evolution of the long-lived component decay time $\tau_{\rm L}$ as a function of energy for temperatures ranging between \SI{5} and \SI{300}{\kelvin}. Decay times extracted from $\gsl^{(2)}(\tau)$ traces measured on QD$_{\rm A}$ at \SI{5}{\kelvin} and at RT are also reported for comparison (blue and red stars). The solid line serves as a guide to the eye.
	} 
	\label{figTRPL5:TemperatureLifetime}
\end{figure}

As shown beforehand (cf. Fig. \ref{figTRPL4:TemperatureDecays}), the recombination kinetics is essentially unaffected when increasing the temperature, which leads to nearly constant exciton lifetimes up to \SI{300}{\kelvin}, regardless of the emission energy (Fig.~\ref{figTRPL5:TemperatureLifetime}). However, the same statement does not apply to the decay time extracted from $\gsl^{(2)}(\tau)$ traces obtained from QD$_{\rm{A}}$, which drops from \SI{16}{\ns} at \SI{5}{\kelvin} to \SI{3.3}{\ns} at RT (Fig. \ref{fig3:Tseries}). While this variation could be ascribed to the difference in oscillator strength between B$_1$ and B$_2$, the drop in $\tau_{\rm decay}$ observed between 150 and \SI{300}{\kelvin} --whereas excitonic recombination is supposed to originate from the B$_{2}$ state at these two temperatures-- is in clear contradiction with TRPL trends. Part of the explanation could lie in the cw quasi-resonant excitation scheme used for auto-correlation measurements, in that the resonant or near-resonant pumping of an exciton excited level may alter the second order auto-correlation function. In this latter case, the characteristic antibunching time could be driven by both the exciton radiative lifetime and the coherence time \cite{Nguyen2011}. However, a quantitative assessment of this latter component would involve challenging experiments in the UV spectral range such as time-integrated four-wave mixing measurements in order to estimate the exact contribution of pure dephasing processes \cite{Borri2001}.

At this stage, we are still left with the fact that the $\gsl^{(2)}(\tau)$ decay time deduced at \SI{300}{\kelvin} matches that of the long-lived component of the TRPL transients, $\tau_{\rm L}$ (cf. Fig.~\ref{figTRPL5:TemperatureLifetime}). Unfortunately, this is in disagreement with the former attribution of $\tau_{\rm L}$ to recombination originating from the low-energy exciton state, the RT $\gsl^{(2)}(\tau)$ trace being issued from the X$_2$ line.

The reported thermal insensitivity of the long-lived component of TRPL transients raises an additional question about the temperature dependence of the emission features of GaN/AlN QDs. Indeed, thermal expansion of the dot and matrix materials is known to reduce their bandgap, and incidentally the QD emission energy. Hence, TRPL transients collected within a fixed energy window are supposed to originate from different QDs when increasing the temperature. We first estimated the energy shift by monitoring the change in the PL peak energy of the QD ensemble. We surprisingly observed a limited redshift of less than \SI{10}{\milli\ev}, which diverges from the $\sim\SI{60}{\milli\ev}$ redshift formerly reported on a GaN/AlN QD ensemble peaking around \SI{4.4}{\ev} \cite{Renard2009shift}. On the other hand, we also had access via \mupl measurements to the thermal shift experienced by QD$_{\rm A}$, which amounts to \SI{60}{\milli\ev}. This value differs from the $\sim \SI{90}{\milli\ev}$ redshift reported in previous works for single GaN QDs emitting at a similar energy \cite{Holmes2014,Tamariz2020}. Hence, it appears that instead of following a universal trend, the emission energy shift of each individual QD upon increasing temperature will likely depend on fluctuations, e.g., in their local strain field. Given the small variations in exciton lifetimes observed for high energy QDs ($> \SI{3.8}{\ev}$), thermal energy shifts of tens of \SI{}{\milli\ev} should not impact their recombination dynamics to a level that can be detected in our experiments. As far as QDs emitting below \SI{3.8}{\ev} are concerned, they display an increasing decay time, which roughly doubles every \SI{200}{\milli\ev} as seen in Fig.~\ref{figTRPL5:TemperatureLifetime}. However, given the small thermal shift undergone by these large QDs, as can be verified once more from the marginal shift reported for the PL peak energy of the QD ensemble for dots emitting near 3.5 eV (cf. Fig. \ref{figTRPL1:Energy}\textbf{a}), the related exciton lifetime variation upon increasing temperature is also expected to remain undetected.\newline

\begin{figure}
	\centering
	\includegraphics[width= 0.48\textwidth]{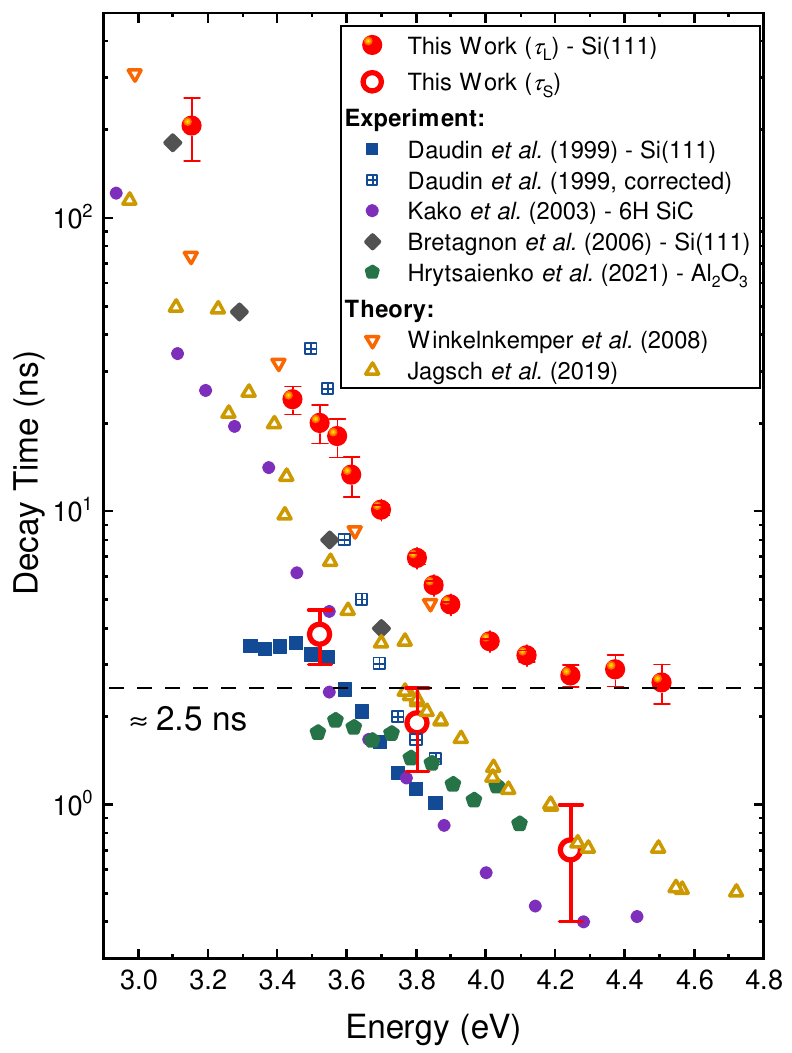}
	\caption{Comparison of exciton decay times reported in GaN/AlN QDs at cryogenic temperature. Experimental results \cite{Daudin1999,Kako2003a,Bretagnon2006,Hrytsaienko2021} are highlighted with filled markers and the substrate employed by each group is specified in the legend.  Computed lifetimes (empty up and down triangles) \cite{Winkelnkemper2008,Jagsch2019} were obtained for various aspect ratios and QD heights. Crossed squares correspond to experimental results after subtraction of a nonradiative recombination component, as detailed in Ref.~\cite{Daudin1999}.
} 
	\label{figTRPL3:Summary}
\end{figure}

\subsection*{Review of excitonic decay in GaN/AlN SA QDs\label{Subsec2.6:TRPLreview}}

To complete our analysis of the exciton recombination dynamics, we compare the long-lived decay times $\tau_{\rm L}$ extracted at \SI{5}{\kelvin} with experimental and theoretical exciton lifetimes published in the literature on SA GaN/AlN QDs \cite{Daudin1999,Kako2003a,Bretagnon2006,Hrytsaienko2021}. In addition, the few short-lived decay times $\tau_{\rm S}$ extracted from power-dependent TRPL experiments have also been added to the picture and all the results are shown in Fig.~\ref{figTRPL3:Summary}. The energy-dependent evolution of $\tau_{\rm L}$ is in line with previous observations: an exponential decrease with decreasing dot size is observed for QDs emitting between 3 and \SI{4.2}{\ev}, followed by a saturation for higher energies. This trend is mainly driven by the electron-hole wave function overlap along the \textit{c}-axis and can be satisfactorily reproduced by approximating the QDs as a two-dimensional quantum well subjected to a built-in electric field \cite{Hrytsaienko2021}.
The same behavior is qualitatively observed for the short-lived component with $\tau_{\rm L}/\tau_{\rm S} \simeq 5$. $\tau_{\rm L}$ matches radiative lifetimes reported earlier on large QDs by Bretagnon \textit{et al.} \cite{Bretagnon2006}. For small QDs, however, exciton lifetimes reported by Kako \textit{et al.} \cite{Kako2003a} and Hrytsaienko \textit{et al.} \cite{Hrytsaienko2021} are closer to the short lifetimes $\tau_{\rm S}$. Let us note that these two groups additionally observed a longer-lived tail in the TRPL transients beyond exciton recombination, which resembles the bi-exponential decay we report here. These recombination processes were ascribed to QD refilling via interface related traps or free carriers present in the WL, and dark states acting as a reservoir. However, given the temperature-independence of $\tau_{\rm L}$ in our case and the quasi-resonant excitation scheme we used, these hypotheses cannot account for our results. Besides, the long-lived decay we observe occurs on a too short timescale compared to that reported in Ref.~\cite{Hrytsaienko2021} and the long-lived component reported in this latter work does not change with emission energy, hence hinting at different mechanisms.

In essence, the data reported in Fig.~\ref{figTRPL3:Summary} illustrate the variability in the exciton radiative recombination lifetime in SA GaN/AlN QDs.
We can still distinguish between two QD excitation regimes: for large QDs, emitting below $\sim\SI{3.5}{\ev}$, the in-plane electron-hole wave function confinement has a negligible impact on the QD PL properties. Hence, to a given QD height is associated a single QD energy--exciton lifetime combination. In other words, as for quantum wells, the QD properties are essentially dominated by the vertical confinement experienced by electron-hole pairs.  For smaller QDs, on the other hand, lateral confinement is expected to significantly impact the QD emission energy \cite{Hrytsaienko2021}. However, fluctuations in the aspect ratio cannot solely account for exciton lifetimes differing by up to one order of magnitude for those emitting above \SI{4}{\ev}. Alternatively, the relief of tensile strain in the AlN matrix induced by microcracks has been shown to increase the excitonic lifetimes of SA GaN/AlN QDs grown on Si(111) substrates \cite{Sarusi2007}. Thus, we reckon that the spread in experimental results is also likely caused by variations in the different strain field applied on QDs grown under distinct conditions. Despite the partial strain relaxation of the AlN matrix after growing a few monolayers, a residual stress originating from the substrate still remains. Thus, for our GaN QDs grown on Si(111) substrate the AlN matrix is still subjected to some biaxial tensile strain \cite{Sarusi2007}. In this regard, we point out that the same substrate was used by Bretagnon \textit{et al.} \cite{Bretagnon2006} for the growth of their dots, whose measured exciton decay times are the closest to ours. Shorter lifetimes have been reported on samples grown on \textit{c}-plane sapphire \cite{Daudin1999,Hrytsaienko2021} and 6H-SiC substrates \cite{Kako2003a} for which the AlN matrix is largely relaxed \cite{Ponce1994,Ponce1995}. Eventually, the QD optical properties will also depend on how the strain of the AlN matrix is impacting the GaN dot morphology (interatomic spacing, etc.). Our QDs are characterized by an $\sim 1.5$ monolayer thick WL, about twice thinner than the one reported in earlier studies \cite{Daudin1999,Kako2003a,Bretagnon2006,Hrytsaienko2021}. Hence, we can expect the AlN matrix to exert a stronger strain on our QDs than in other samples. Besides, a thinner WL should result in a stronger electronic confinement of the electrons and holes trapped in the QDs \cite{Sanguinetti2003,Sun2012}, impacting both the exciton lifetime and the emission energy. In a pyramidal metal-polar GaN QD, however, the hole state is strongly confined due to its large effective mass and the electron is localized at the top of the dot. Hence, the impact of the WL thickness on their respective wave functions may be limited.

In summary, changes in the strain state from one type of sample to the other related to the nature of the substrates and the thickness of the WL could account for the reported variations in the exciton lifetime. In this regard, systematic micro-Raman spectroscopy measurements could be a complimentary and insightful technique to draw correlations between the strain state of QD ensembles and the measured lifetimes.

\section*{Discussion\label{Sec3:discussion}}

In conclusion, we have studied the evolution of the SPE behavior of polar SA GaN/AlN QDs from \SI{5}{\kelvin} up to RT and provided a detailed description of the framework used to analyze their luminescence properties. We have reported single photon purity values as low as $g^{(2)}(0) = \SI{0.05(2)}{}$ at \SI{5}{\kelvin} and $\SI{0.17(8)}{}$ at \SI{300}{\kelvin} by taking advantage of the large spectral separation between exciton and biexciton emission lines on the one hand and the quasi-resonant excitation scheme on the other hand. We have subsequently investigated the recombination dynamics of excitons for energies ranging from 3.2 to \SI{4.5}{\ev}, i.e., for QDs that emit both below and above the bulk GaN bandgap. This allowed us to evidence a two-component recombination process that persists in the low power density regime for any QD emission energies. Both fast and slow decay rate components are observed to drop significantly with increasing QD size. This follows the characteristic increase in exciton lifetime related to decreasing electron-hole wave function overlap along the vertical \textit{c}-axis in polar quantum heterostructures. The long-lived exponential decay time remains constant, within experimental uncertainties, from cryogenic temperatures up to \SI{300}{\kelvin}. This allowed us to associate it with a radiative recombination process stemming from an exciton bright state. This temperature invariance of the QD photoluminescence decay discards the contribution of a long-lived dark state reservoir to the bi-exponential dynamics, in contrast with earlier results obtained on CdSe QDs \cite{Labeau2003,Patton2003,Sallen2009}. The latter result also confirms the negligible influence of nonradiative processes on the exciton recombination dynamics of GaN/AlN QDs, regardless of their size. As such, these QDs truly emerge as a robust platform for RT quantum applications, where the single photon purity of alternative epitaxial QD emitters breaks down \cite{Yu2019}.

\section*{Materials and methods \label{Sec4:materials}}

\begin{widetext}

\begin{figure}[h]
	\centering
	\includegraphics[width= \textwidth]{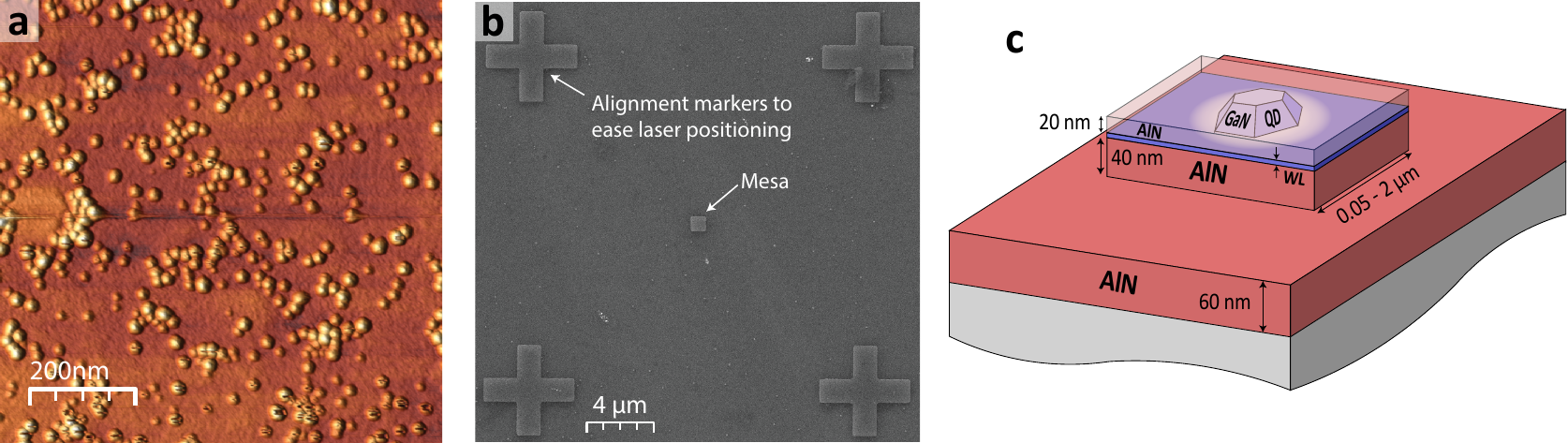}
	\caption{\textbf{a.} $1 \times 1 \ \si{\square\mum}$ AFM image taken in height mode of the top plane of GaN QDs  displaying a broad distribution of dot sizes. \textbf{b.} $26 \times 26 \ \si{\square\mum}$ secondary electron image of the SA QD sample after evaporation of the top QD plane and etching into mesa structures. The displayed mesa is $1 \ \si{\mum}$ wide. 
	\textbf{c.} Sketch of the mesa structure. In practice, several QDs can be present on
each mesa depending upon their dimension.
	 }
	\label{figLAST:materials}
\end{figure}

\end{widetext}

The wurtzite SA QD sample under scrutiny was grown by ammonia-source molecular beam epitaxy (NH$_{3}$-MBE). It first consists of a 100-nm-thick metal-polar AlN layer deposited on a Si(111) substrate. Then, a plane of SA GaN QDs ($\sim$5-monolayer-thick) was grown and capped by a 20-nm-thick AlN barrier. At the surface, an additional uncapped plane of QDs was grown to check their size and density using atomic force microscopy (AFM) imaging (Fig.~\ref{figLAST:materials}\textbf{a}). This latter QD plane was then evaporated under vacuum in the MBE chamber to guarantee the integrity of the remaining QD layer. The WL emission was detected by cathodoluminescence at \SI{5.3}{\ev} at a temperature of \SI{12}{\kelvin}, which corresponds to a GaN WL thickness of $\sim$1.5 monolayers \cite{Liu2018}. The sample was then patterned into mesas scaling from $ 0.05 \times 0.05 \ \si{\square\mum}$ up to $ 2 \times 2 \ \si{\square\mum}$ by electron beam lithography and subsequent etching down to the AlN buffer layer. Figure~\ref{figLAST:materials}\textbf{b} presents a top view of the processed sample obtained by scanning electron microscopy and the mesa structure is schematized in Fig.~\ref{figLAST:materials}\textbf{c}. Further details about SA QD growth can be found in Tamariz \textit{et al.} \cite{Tamariz2019}.

\mupl and second order correlation function measurements were performed under quasi-resonant excitation by generating electron-hole pairs in the QD excited states. To this aim, we employed two different continuous wave (cw) laser light sources emitting either at \SI{266}{\nm} (\SI{4.66}{\ev})  or at \SI{244}{\nm} (\SI{5.08}{\ev}), i.e., below the WL transition energy. We collected $\gsl^{(2)}(\tau)$ traces using an HBT interferometer with a time resolution of \SI{220}{\pico\second}.  We recorded TRPL transients using a pulsed \SI{266}{\nm} laser with an \SI{8.45}{\kilo\hertz} repetition rate and a pulse duration of \SI{440}{\pico\second}, which is shorter than the GaN/AlN QD decay rate. The experimental setup is further detailed in the SI Sec. [S1].

{\footnotesize
\subsection*{Acknowledgements}

This work was supported by the Swiss National Science Foundation through Grants 200021E\_15468 and 200020\_162657 and by the Marie Sklodowska-Curie action "PhotoHeatEffect" (Grant No. 749565) within the European Union's Horizon 2020 research and innovation program.

\subsection*{Conflict of interests}

The authors declare no competing interests.

\subsection*{Contributions}
 
S.T. grew the samples. J.S, S.T. and G.C. performed the \mupl and TRPL measurements. J.S, S.T., G.C. and R.B. analyzed the data. J. S. and R.B wrote the manuscript. N.G. initiated and supervised the entire project. All authors contributed to the discussion of the results.
}

{\color{white}0

0

0}

\bibliography{Light}

\end{document}


\title{Supplementary Information: Single photon emission and recombination dynamics in self-assembled GaN/AlN quantum dots}

\author{Johann Stachurski}
\email{johann.stachurski@epfl.ch}
\affiliation{Institute of Physics, \'Ecole Polytechnique F\'ed\'erale de Lausanne, EPFL, CH-1015 Lausanne, Switzerland}
\author{Sebastian Tamariz}
\affiliation{Institute of Physics, \'Ecole Polytechnique F\'ed\'erale de Lausanne, EPFL, CH-1015 Lausanne, Switzerland}
\affiliation{Current address: Université Côte d’Azur, CNRS, CRHEA, F-06560 Valbonne, France}
\author{Gordon Callsen}
\affiliation{Institute of Physics, \'Ecole Polytechnique F\'ed\'erale de Lausanne, EPFL, CH-1015 Lausanne, Switzerland}
\affiliation{Current address: Institut für Festkörperphysik, Universität Bremen, 28359 Bremen, Germany}
\author{Rapha\"el Butt\'e}
\affiliation{Institute of Physics, \'Ecole Polytechnique F\'ed\'erale de Lausanne, EPFL, CH-1015 Lausanne, Switzerland}
\author{Nicolas Grandjean}
\affiliation{Institute of Physics, \'Ecole Polytechnique F\'ed\'erale de Lausanne, EPFL, CH-1015 Lausanne, Switzerland}

\maketitle

\newpage

\beginsupplement

\section{\mupl and TRPL optical characterization methods}

\begin{figure*}
	\centering
	\includegraphics[width= 1\textwidth]{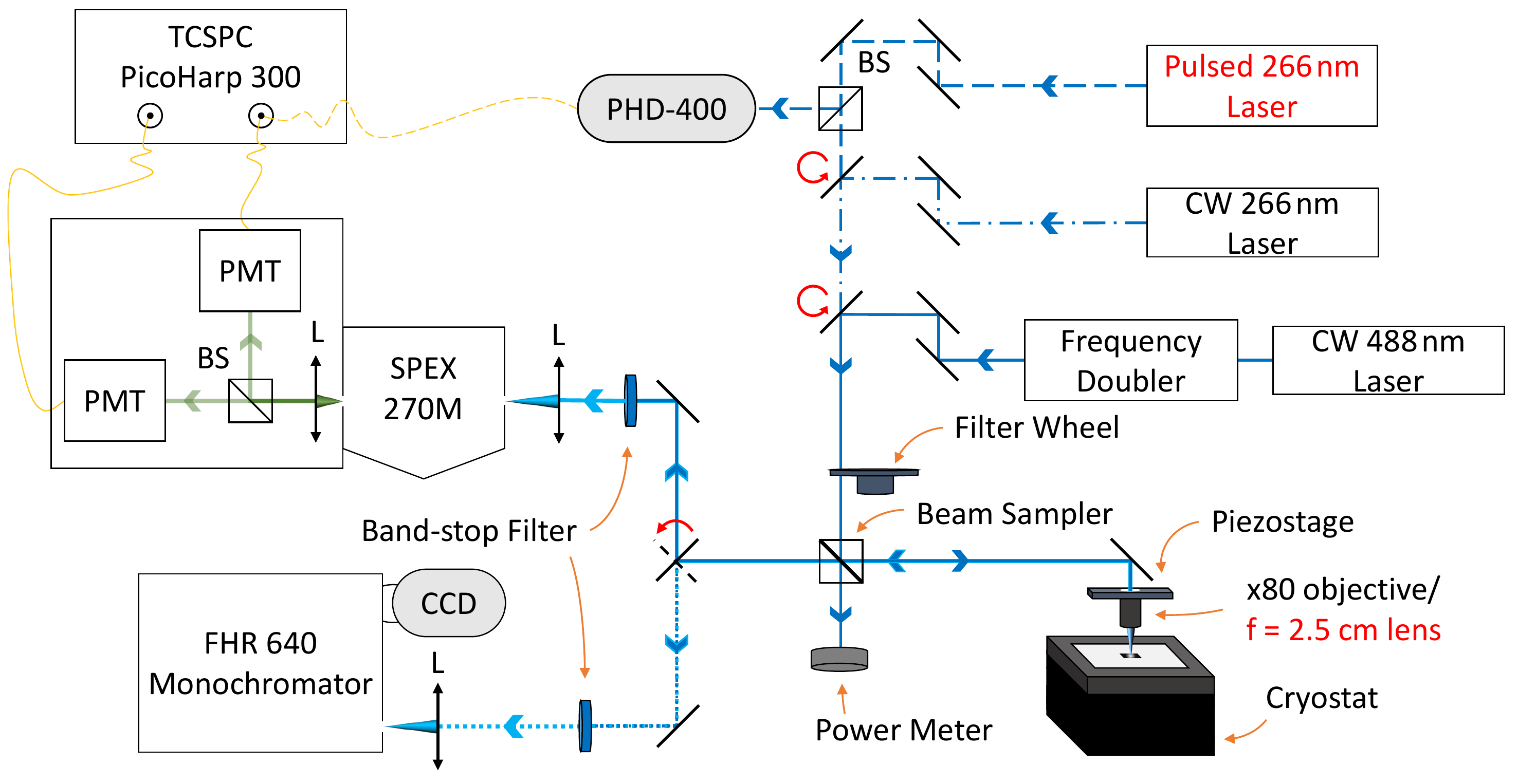}
	\caption{Schematic illustration of the optical setup. BS and L stand for beam splitter and lens, respectively. Dashed, dotted and dash-dotted lines correspond to alternative optical paths. The red arrows indicate flippable mirrors.
	}
	\label{figS2:setup}
\end{figure*}

\begin{figure*}
	\centering
	\includegraphics[width= 0.8\textwidth]{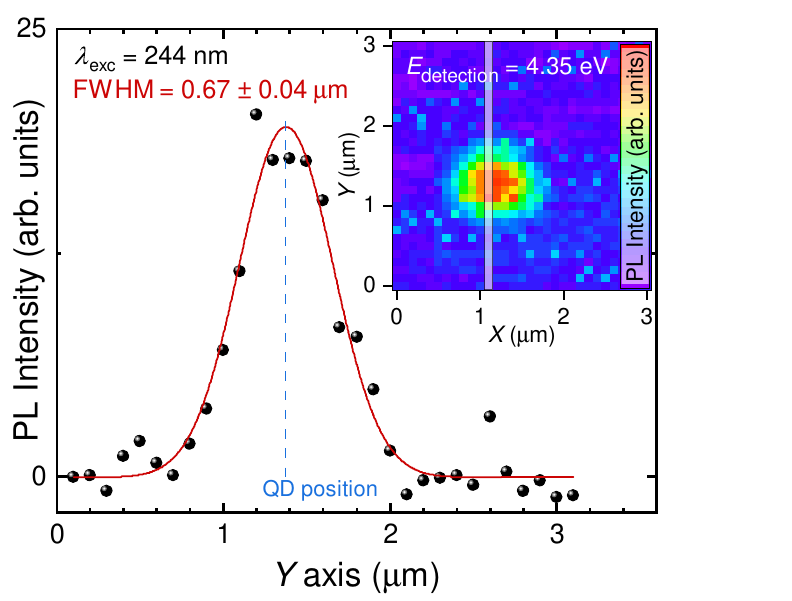}
	\caption{ Gaussian profile of the \mupl \SI{244}{\nm} laser spot recorded by scanning the PL emission of a single QD emitting at $E=\SI{4.35}{\eV}$. Inset: mapscan centered on the QD. The PL profile corresponds to the points encompassed into the white stripe. The laser spot displays a Gaussian profile with a full width at half-maximum (FWHM) of \SI{0.67}{\mum}.
	 }
	\label{figS1:Resolution}
\end{figure*}

Microphotoluminescence (\textmu-PL) and second order auto-correlation (\gsl$^{(2)}(\tau)$) measurements were performed under continuous-wave (cw) excitation using either a \SI{266}{\nm} (\SI{4.66}{\ev}) Nd:YAG laser from Crylas or a \SI{488}{\nm} solid state laser (Coherent Genesis CX SLM) frequency-doubled to \SI{244}{\nm} (\SI{5.08}{\ev}) by a Spectra-Physics WaveTrain. For time-resolved photoluminescence (TRPL) measurements, we used a Teem Photonics Microchip \SI{266}{\nm} pulsed Nd:YAG laser with a repetition rate of \SI{8.45}{\kilo\hertz} and a pulse duration of 440 ps. The different sources are depicted on the illustration of the setup shown in Fig.~\ref{figS2:setup}.
Under cw excitation, we focused the laser beam with a $\times$80 Mitutoyo microscope objective (${\rm NA} = 0.55$) to reach a spot size of $\simeq \SI{1}{\micro\meter}$ (Fig.~\ref{figS1:Resolution}).

For TRPL measurements, the laser was focused using a conventional plano-convex lens of \SI{2.5}{\centi\meter} focal length to reach a spot size of $\simeq \SI{100}{\mum}$. In order to optimize the collection of the sample photoluminescence, we guided each laser beam using a 90:10 transmission/reflection beam sampler onto the sample, which was placed in a closed-cycle helium cryostat (Cryostation C2 from Montana Instruments, Inc.) to cool it down to \SI{5}{\kelvin}. \mupl mapping of the surface was performed by mounting the microscope objective on a Physik Instrumente P-612.2 XY piezostage. \mupl spectra were recorded with a Horiba Symphony II UV-enhanced charge-coupled device (CCD) coupled to a Horiba FHR 640 monochromator equipped with  1800 l/mm and 150 l/mm holographic gratings. Autocorrelation measurements were performed using a Hanbury-Brown and Twiss (HBT) interferometer with two PicoQuant PMA 175 photomultiplier tubes linked to a PicoHarp 300 time-correlated single photon counter (TCSPC).  For TPRL measurements, a PHD-400 photodiode module was alternatively connected to one channel of the TCSPC to use the laser pulses as a trigger. We measured a time resolution of \SI{220}{\pico\second} for the autocorrelation setup. The bandpass for \gsl$^{(2)}$ and TRPL measurements was selected using a SPEX 270M monochromator with a 2400l/mm grating and a slit opening of \SI{1.6}{\milli\meter}. The collection efficiency of the HBT setup was estimated to \SI{0.27(10)}{\percent}. Further details can be found in Tamariz \textit{et al.} \cite{Tamariz2020}.

\section{Photon extraction efficiency}

Besides the HBT setup collection efficiency reported above, we should ideally compute the photon extraction efficiency from our sample to relate the detection rate to the real QD emission rate. While an exact computation of this latter efficiency goes beyond the scope of this work given the complexity introduced by the mesa geometry, we can nonetheless give a lower bound value for the extraction efficiency $\eta_{\rm ext,min}$ by approximating our system to a  conventional planar geometry while only considering photons emitted toward the top AlN/air interface, i.e., the photon flux partly reflected at the AlN/silicon substrate interface is neglected. Within this framework, the QD is considered as an isotropic emitter and the optical losses are only assumed to occur at the AlN/air interface. Owing to its wide bandgap, the AlN layer is fully  transparent in the whole domain of emission of the QDs. The transmission coefficient $T$ under normal incidence, also known as the dielectric efficiency \cite{Rosencher2002}, obtained using the refractive index of AlN $n_{\rm AlN}\approx 2.13$ \cite{Brunner1997} around \SI{4.5}{\ev}, is defined as:
\begin{equation}
    \label{eq:extraction} T = 1 - \frac{(n_{\rm AlN}-1)^2}{(n_{\rm AlN}+1)^2} = 0.87.
\end{equation}

$\eta_{\rm ext,min}$ depends on the solid angle leading to light extraction $\Omega_c$, which simply writes:
\begin{equation}
   \label{n_extr} \Omega_c = \frac{1}{2}(1-\cos(\theta_c))\cdot \Omega_{tot},
\end{equation}
where $\theta_c = \arcsin(n_{\rm AlN}^{-1})$ is the critical angle leading to total internal reflection and $\Omega_{\rm tot} = 4\pi$ is the total solid angle. As a result, we get $\eta_{\rm ext,min} \approx T \cdot \Omega_c/\Omega_{\rm tot}$, which leads to $\eta_{\rm ext,min} =$ \SI{5.1}{\percent}. In pratice, we expect the actual extraction efficiency to be significantly higher. Note also that given the much smaller refractive index value of wide bandgap semiconductors compared to their III-arsenide counterparts, the present $\eta_{\rm ext,min}$ value is more than three times larger than that we would obtain for InAs/GaAs QDs.

\section{Polarization- and temperature-dependent QD photoluminescence}

\begin{figure*}
	\centering
	\includegraphics[width= 0.8\textwidth]{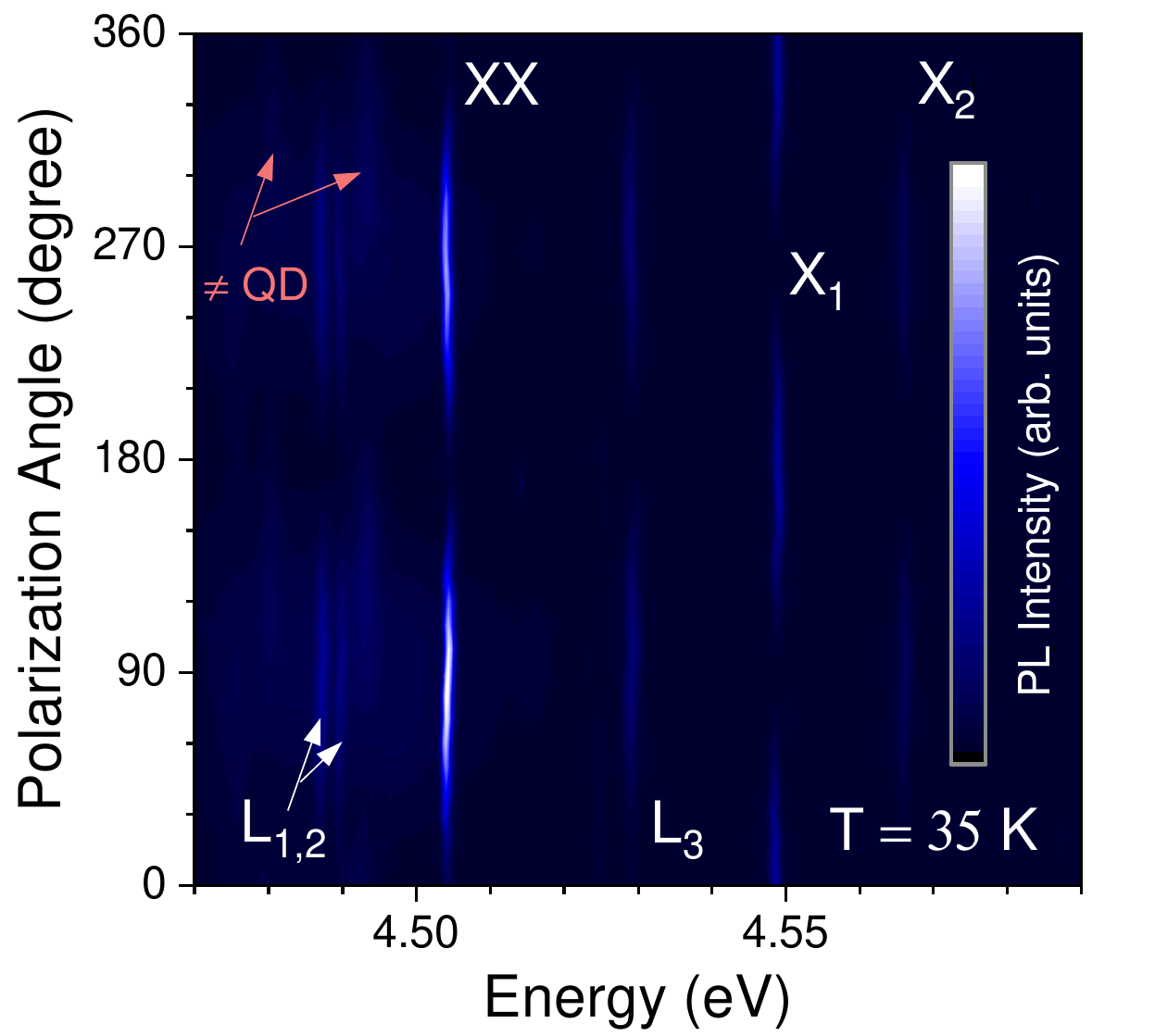}
	\caption{Color map of polarization-dependent photoluminescence spectra of QD$_{\rm A}$ recorded at $T=$ 35 K. The lines L$_1$ to L$_3$ have not been clearly identified yet. The low energy lines are overlapping with the emission of another QD. Data taken under cw excitation at a wavelength of \SI{244}{\nano\meter} and a power density of \SI{400}{\kilo\watt\per\square\centi\meter}.
	}
	\label{figS3:Polarization}
\end{figure*}

Here, we present additional photoluminescence measurements, mostly performed on the quantum dot (QD) labelled QD$_{\rm A}$. This QD is especially interesting for its brightness and the large spectral separation between its emission lines. Additional peaks are however observed in the same spectral window and are attributed to another QD based on their different linear polarization orientation  (Fig.\ \ref{figS3:Polarization}). The two QD$_{\rm A}$ biexciton (XX) lines are not resolved, while the two exciton lines (X$_{1,2}$) are fully separated. The lines L$_1$ to L$_3$ have not been assigned so far, but they show the same linear polarization orientation than XX and X$_2$. X$_1$ is cross-polarized to all other lines of QD$_{\rm A}$. Apart from L$_3$, similar patterns have been observed earlier on GaN/AlN QDs \cite{Honig2014,Arita2017,Tamariz2020}.

\begin{figure*}
	\centering
	\includegraphics[width= 0.8\textwidth]{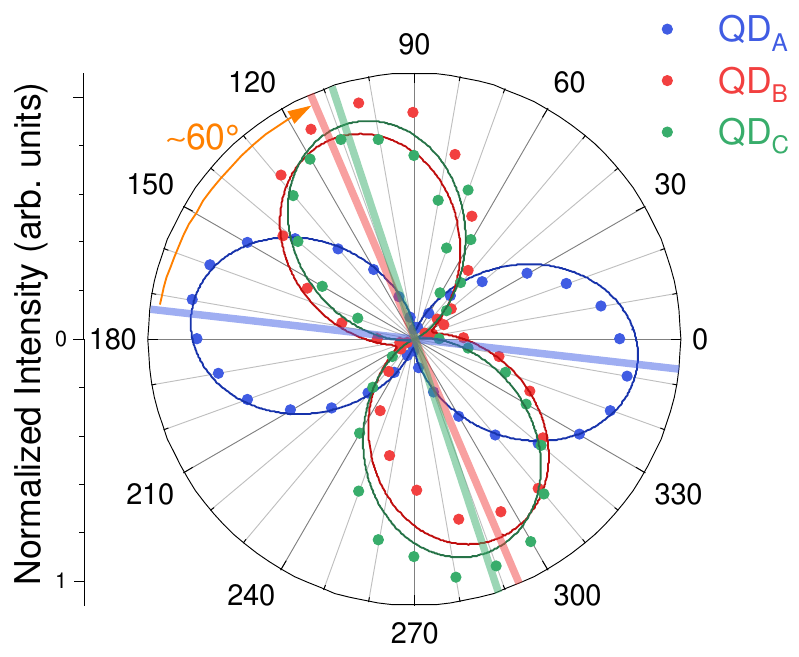}
	\caption{Polar representation of the normalized intensity of the $X_1$ line measured for three different QDs. Solid lines are fitted using Eq. (\ref{eq:pola}). All excitonic lines display a strong degree of linear polarization ($\rho>0.95$) with polarization axes highlighted with thick solid lines on the figure. Polarization axes of QD$_{\rm A}$-X$_1$ and QD$_{\rm A}$-X$_2$ are rotated by \SI{61(2)}{\degree}.}
	\label{figSb1:polaQDs}
\end{figure*}

The comparison between the polarization-dependent photoluminescence measurements of different QDs performed at cryogenic temperature suggests that the polarization of exciton emission lines could be pinned by crystallographic axes. As an illustration, Fig.~\ref{figSb1:polaQDs} shows the polar representation of the normalized intensity of three X$_1$ exciton emission lines. The data are well reproduced when using the relationship \cite{Bardoux2008}:
\begin{equation} \label{eq:pola}
  I(\theta) = a + b \cdot\cos{(\theta-\theta_0)}^2,  
\end{equation}

where $a, b$ and $\theta_0$ are fitting parameters. The linear polarization degree $\rho = (I_{\rm max}-I_{\rm min})/(I_{\rm max}+I_{\rm min})$, where $I_{\rm max}$ and $I_{\rm min}$ are the maximum and minimum X$_{1}$ line integrated $\mu-$PL intensity, respectively,  is systematically above \SI{95}{\percent}.  The polarization axis of the QD$_{\rm A}$-X$_1$ line is tilted by about \SI{60}{\degree} with respect to QD$_{\rm B,C}$-X$_1$, hence recalling the in-plane hexagonal symmetry of wurtzite GaN. However, we do not have a statistically significant set of data to fully validate this preliminary result. A previous report on the polarized emission of single GaN/AlN QDs by Bardoux and co-workers \cite{Bardoux2008} did not support the existence of such preferential polarization axes while no unequivocal conclusion arose from the results reported in the PhD thesis by Kindel \cite{Kindel2010a}.

\begin{figure*}
	\centering
	\includegraphics[width= 0.8\textwidth]{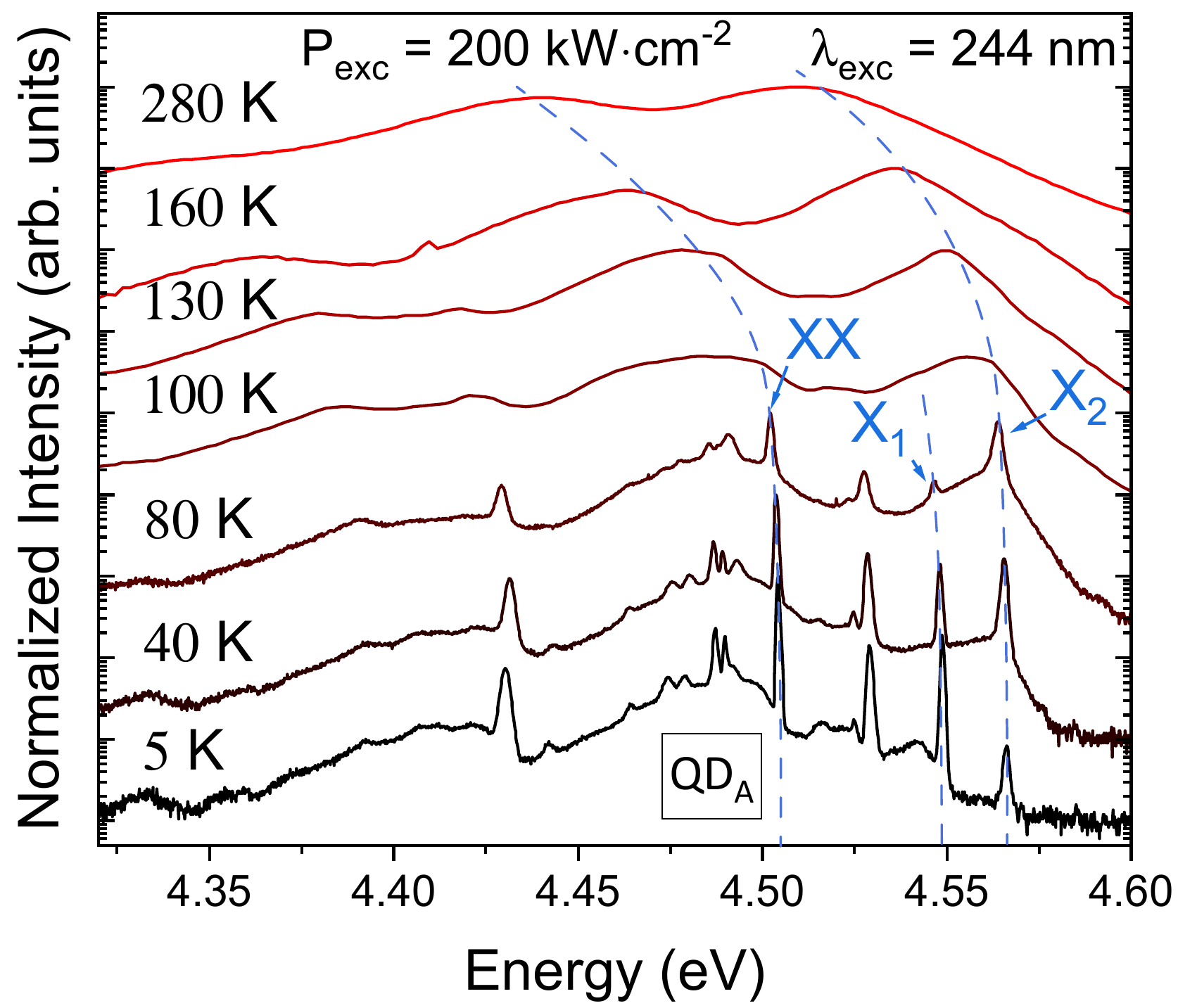}
	\caption{Temperature-dependent \mupl spectra recorded on QD$_{\rm A}$. The exciton and biexciton lines are labelled. The dashed lines serve as a guide to the eye. X$_2$ takes over X$_1$ around \SI{40}{\kelvin}. The low energy lines are overlapping with the emission of another QD. The smoothening observed between \SI{80}{\kelvin} and \SI{100}{\kelvin} spectra stems from a grating change. Low (up to $T=$ 80 K) and high temperature spectra were recorded with a 1800 l.mm$^{-1}$  and a 150 l.mm$^{-1}$ grating, respectively.
	 }
	\label{figS4:Temperature}
\end{figure*}

At \SI{5}{\kelvin}, XX and X$_1$ dominate the emission spectrum. The transition from dark to bright states requires the simultaneous absorption of phonons which match their energy difference \cite{Honig2014}. Assuming a negligible dark-state splitting, the intensity of X$_2$ increases when the thermal energy $k_BT$ approaches $E_{\rm{DB}} + E_{\rm{BB}}/2$, where $E_{\rm{BB}}$ ($E_{\rm{DB}}$) is the bright (dark-to-bright) state splitting. As observed in Fig.\ \ref{figS4:Temperature}, X$_2$ takes over X$_1$ in intensity around \SI{40}{\kelvin}.

\section{Limitations of the second order auto-correlation fitting function and impact of charge fluctuations}

The standard multi-level model (Eq.~1 of the main text) used to fit the second order autocorrelation function measurements predicts a narrowing of the antibunching dip as a consequence of a bunching behavior. This arises from an increase in the QD mean occupation number $\mu = \Pi/\gamma$, where the decay rate $\gamma = 1/\tau_{\rm decay}$, with $\tau_{\rm decay}$ the exciton decay time, is assumed fixed and the pump rate $\Pi$ depends on the excitation power density. This narrowing is visible in Fig.~2\textbf{c} of the main text. However, the fit partially diverges from the data when $\gamma$ is kept constant. A more consistent fit can only be obtained when assuming an increase in $\gamma$ with excitation power density. Such a fit yields a $\tau_{\rm decay}$ value that drops from about 19 to \SI{13}{\ns} between 40 and \SI{400}{\kilo\wcm}. This variation may be accounted for by different phenomena that add to the contribution of dark states to the exciton dynamics discussed in the main text. These alternative explanations are discussed hereafter.

In polar III-nitride QDs, the strong internal electric field $\textbf{E}_{\rm in}$ due to the macroscopic polarization mismatch between the dot and the barrier materials is responsible for the out-of-plane separation of electron and hole wave functions. This separation modifies the oscillator strength of the exciton ground state, which in turn drives the variations in the decay rate with QD size. A variation in $\textbf{E}_{\rm in}$ with increasing excitation power density could therefore possibly explain the change in $\gamma$ observed via the correlation measurements shown in Fig.~2\textbf{c} of the main text. Screening of the built-in field due to trapping of additional carriers at high excitation power densities has already been reported in GaN/AlN QDs \cite{Bretagnon2006} and may, at first sight, explain this variation. At the level of a single QD, however, this results in sharp changes in the QD emission energy and in the observation of additional emission lines (biexciton, charged exciton, etc.), which do not contribute to the $\gsl^{(2)}(\tau)$ traces. Hence, the fluctuations of $\textbf{E}_{\rm in}$ can only be accounted for by charge fluctuations external to the QDs. A common way to picture it consists in considering the charge and discharge of defect states (DSs), whose origin could be ascribed to the presence of vacancies at the GaN/AlN interface, surface states, or point defects surrounding the QD. First, setting aside their exact origin, DSs are responsible for a time-dependent extrinsic electric field $\textbf{E}_{\rm ext}(t)$, which creates an additional Stark shift of the exciton emission energy given by
\begin{equation}\label{eq1:deltaE}
\Delta E(t) = \bm{\mu} \cdot \textbf{E}_{\rm ext}(t),
\end{equation}
where $\bm{\mu}$ is the exciton dipole moment, which is aligned along the $c$-axis and is oriented toward the top of the QD in metal-polar III-nitride QDs. DSs can be modelled as traps with given capture $\tau_{\downarrow}$ and escape $\tau_{\uparrow}$ times. At cryogenic temperature and under low excitation conditions, the phonon-mediated escape of trapped carriers is blocked and DSs are assumed to be mostly charged \cite{Lohner1992,Berthelot2006}. Carrier escape is triggered when the sample is pumped and the higher the excitation power density, the lower the probability for a defect to be occupied. In the extreme situation where all the defects are depleted, the exciton energy gets ultimately shifted by an energy $\Delta E_{\rm max}$, which accounts for the contribution of all DSs. As developed in Ref.~\cite{Kindel2010a}, for DSs in the close vicinity of polar III-nitride QDs, the exciton peaks are expected to shift with increasing excitation power density following

\begin{equation} \label{eq2:Shift}
    E(P_{\rm exc}) = E_0 + \frac{\Delta E_{\rm max}}{\sqrt{P_{\rm exc}/P_{50\%}}+1}.
\end{equation}

In Eq. (\ref{eq2:Shift}), $E_0$ corresponds to the emission energy of the excitonic line of interest when all DSs are filled whereas $P_{50\%}$ corresponds to the excitation power density for which the occupation probability of the DSs is equal to one half. The sign of $\Delta E_{\rm max}$ depends on the overall positioning of the defects around the QDs and on the sign of the trapped carriers.

\begin{figure*}
	\centering
	\includegraphics[width= 0.8\textwidth]{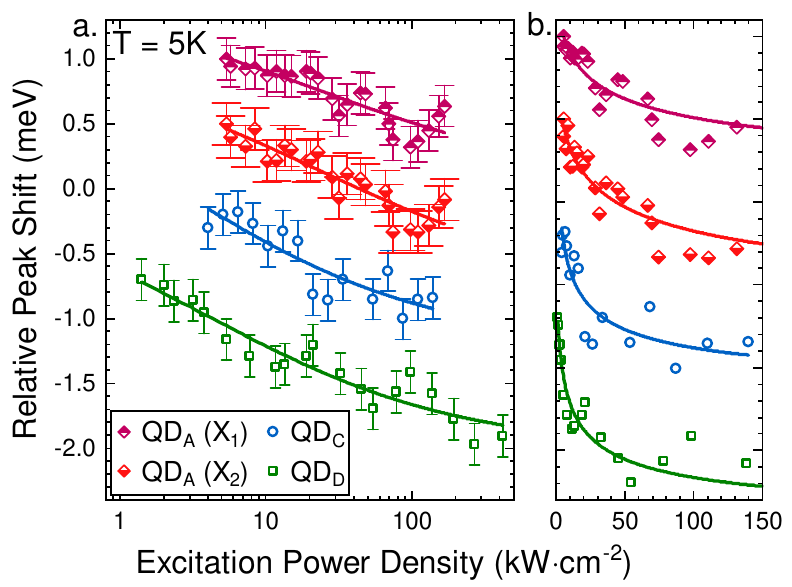}
	\caption{Power-dependent evolution of the exciton peak energy for QDs emitting between 4.4 and \SI{4.6}{\ev}, recorded at \SI{5}{\kelvin} and displayed in \textbf{a.} and \textbf{b.} using logarithmic and linear excitation power density scales, respectively. The results (data points) are fitted  using Eq.~\ref{eq2:Shift} (solid lines). QD$_{\rm A,C}$ were excited using the \SI{244}{\nm} laser line and QD$_{\rm D}$ using the \SI{266}{\nm} laser line. The shift reported for QD$_{\rm C,D}$ is related to the X$_1$ line. An arbitrary energy offset was applied to each dataset for the sake of visual convenience. The magnitude of the error bars is related to the spectral separation between two consecutive pixels since our measurements are limited by the pixel size in the CCD.
	 }
	\label{figS5:RelativeShift}
\end{figure*}

Such a shift has been consistently observed for various QDs, including QD$_{\rm A}$ on which most of the $\gsl^{(2)}(\tau)$ traces considered in this work have been recorded. The corresponding results are thus reported for three different QDs in Fig.~\ref{figS5:RelativeShift}\textbf{a}. All of the investigated exciton emission lines displayed a redshift of about 0.5 to \SI{2}{\milli\ev}. Interestingly, when adopting a linear excitation power density scale to show the same data (Fig. \ref{figS5:RelativeShift}\textbf{b}), it is clearly seen that this shift mainly occurs for excitation power densities below \SI{30}{\kilo\wcm}. The steady redshift hints at a common origin for $\textbf{E}_{\rm ext}$, such as DSs located at the surface of the \SI{20}{\nm} thick AlN capping layer. An accumulation of point defects occurring above or below the GaN wetting layer (WL) could also account for this electric field of extrinsic origin. In this regard, Si impurities diffusing from the Si(111) substrate to the AlN buffer layer, even in minute concentration, or $n$-type O or C impurities are the most likely to be found in our sample as well as hydrogen-related charged defects.

Importantly enough, the observed shifts remain rather small ($\sim$1 meV, i.e., on the order of the FWHM of the X$_{1,2}$ emission lines), which contrasts with the \SI{30}{\percent} reduction in $\tau_{\rm decay}$ extracted from second order autocorrelation measurements when fitting each $\gsl^{(2)}(\tau)$ trace individually. At first sight, such a decrease in $\tau_{\rm decay}$ would be commensurate with a noticeable change in the oscillator strength of the exciton ground state and hence a large spectral shift. Indeed, in the energy range associated to these dots --QD$_{\rm A,C,D}$ all emit above $\SI{4}{\ev}$-- the energy shift leading to a change of \SI{30}{\percent} in $\tau_{\rm decay}$ would amount to hundreds of meV (cf. Fig.~11 of the main text). Thus, any screening of the built-in electric field can be excluded to explain the measured variation in the decay time. Since the amplitude of the electric field generated by DSs is on the order of a few \SI{}{\mega\volt\per\meter} \cite{Kindel2014}, it cannot impact the built-in electric field of about 7-\SI{9}{\mega\volt\per\centi\meter} \cite{Bretagnon2006,Hrytsaienko2021} significantly enough to explain the measured reduction in the decay time.

Besides variations in the built-in electric field experienced by the ground state exciton, another reason for explaining the divergence between the fit with fixed $\gamma$ and experimental $\gsl^{(2)}(\tau)$ data can also be found in the original hypothesis used to develop the multi-level model. Indeed, to allow for a closed analytic form of the second order autocorrelation function in the framework of this model, the recombination rate of an $N$-exciton is assumed to be proportional to the number of electron-hole pairs in the QD, i.e., $\gamma_N = N \cdot \gamma$ \cite{Kindel2010a}. As the excitation power density is increased and the bunching induced by a higher QD mean occupation number is strengthened, this approximation may become more questionable.

\section{Estimation of the defect density through quantum dot emission linewidth statistics}

\begin{figure*}
	\centering
	\includegraphics[width= 0.8\textwidth]{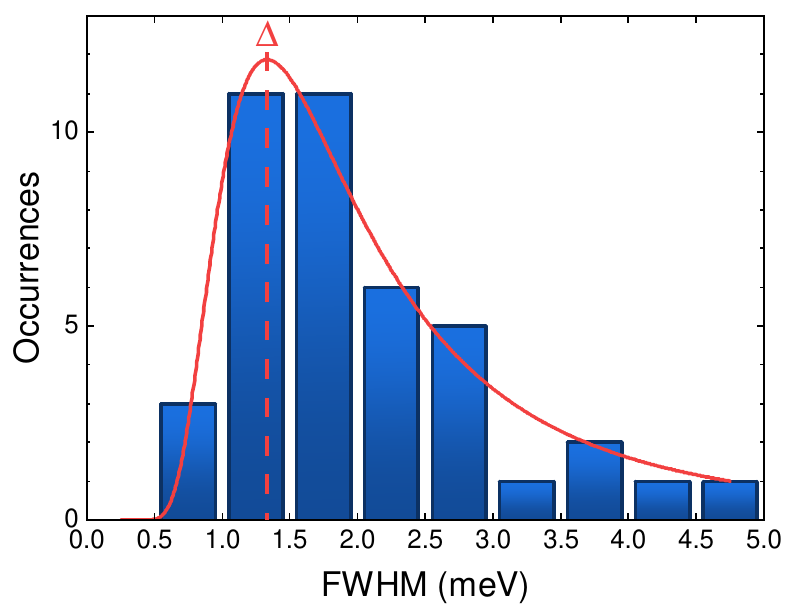}
	\caption{Histogram of the excitonic peak linewidth distribution for QDs emitting between 4.65 and \SI{4.85}{\electronvolt}. The linewidth distribution is fitted with a Fréchet function (Eq.~\ref{eq4:frechet}) times a normalization parameter. The linewidth leading to the maximum value of the distribution, $\Delta = \SI{1.3(3)}{\milli\electronvolt}$, has been determined as an average over various binning conditions.
	 }
	\label{figS6:Frechet}
\end{figure*}

As shown in Ref.~\cite{Kindel2010a}, spectral diffusion in III-nitride polar QDs is also responsible for a very specific excitonic linewidth distribution, which is highlighted in Fig.~1\textbf{d} of the main text. As detailed beforehand, QDs are impacted by the charging and discharging of neighboring DSs. While these charge fluctuations contribute to the broadening of exciton emission lines at cryogenic temperature, way beyond the natural linewidth, this feature can also be used to estimate the defect density in the dot surrounding.

To this end, we relied on the probability distribution function of the exciton linewidth for a given emission energy ($E$) that can be approached by a Fréchet distribution \cite{Kindel2010a}:
\begin{equation}
\label{eq4:frechet} f_{Fr}(E) = \frac{\alpha}{s}\cdot \qty(\frac{E}{s})^{-1-\alpha}\cdot \exp[-\qty(\frac{E}{s})^{-\alpha}],
\end{equation}
where the scaling parameter $s$ and the shape parameter $\alpha$ are considered as free parameters here. The maximum value of $f_{Fr}(E)$ is associated to a linewidth value ($\Delta$) that can then be used to estimate the point defect density $\tilde{\rho}$ using \cite{Kindel2010a}:
\begin{equation}
\label{eq5:density}  \tilde{\rho} = \frac{3}{4\pi}\qty[\frac{1}{2\sqrt{2\ln{2}}}\frac{1}{\sqrt{p{(1-p)}}}]^{3/2}\qty(\frac{\Delta}{\kappa e})^{3/2},
\end{equation}
where $p$ is the occupation probability of a defect, $e$ is the elementary charge, and $\kappa$ is a numerically estimated coefficient that depends on the QD emission energy. Using $p = 1/2$ we can estimate a minimum value for $\tilde{\rho}$. Ideally, $\Delta$ is determined experimentally for a fixed emission energy by collecting the FWHM histogram of exciton emission lines within a narrow enough spectral range and by fitting the results with Eq.~\ref{eq4:frechet}. Depending on the amount of recorded emission lines, this spectral range can be expanded to get a statistically significant dataset. In this work, we extracted the FWHM of all spectrally resolved exciton emission lines following a $100\times\SI{100}{\square\micro\meter}$ PL mapscan collected over a sample area processed into mesa structures. Results are displayed in Fig.~1\textbf{d} of the main text. From this set of measurements, we retained QDs emitting within a \SI{200}{\milli\ev} interval centered at \SI{4.75}{\ev}, which led to the histogram shown in Fig.~\ref{figS6:Frechet}. While the high energy tail of the distribution is not ideally resolved, the number of occurrences remains large enough to estimate the position of the maximum of the distribution, namely $\Delta = \SI{1.3(3)}{\milli\ev}$. The resulting $ \tilde{\rho}$ value, $\sim \SI{1e18}{\per\cubic\centi\meter}$, is a rough estimate of the minimum DS density located in the vicinity of the present polar SA GaN/AlN QDs. In this respect, our value compares well to the $\sim \SI{2e19}{\per\cubic\centi\meter}$ defect density computed in Ref.\ \cite{Kindel2010a} for GaN/AlN QDs grown on SiC.

\section{Impact of the detection window extent on the single-photon purity recorded by second order autocorrelation measurements.}

\begin{figure*}
	\centering
	\includegraphics[width= 0.8\textwidth]{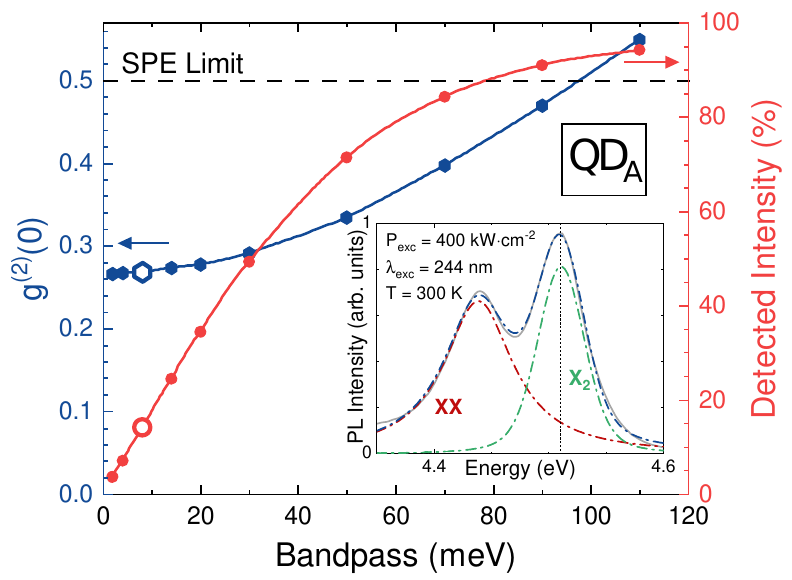}
	\caption{Evolution of single-photon purity and collected exciton PL intensity as a function of the bandpass of the detection window. The detected PL intensity of the exciton line is given as a percentage of the total exciton intensity. The empty hexagon and dot underscore the \SI{8}{\milli\ev} bandpass, which was used experimentally. The horizontal black dashed line outlines the 0.5 single-photon emission limit. Solid lines serve as a guide to the eye and all of the data points were estimated after the two-peak fitting of the QD PL, as depicted in the inset. The blue dashed line is a the sum of a Lorentzian function (XX emission line, red dashed line) and a Voigt profile (X$_2$ emission line, green dashed line) . The black dotted line identifies the center of the detection window.
	 }
	\label{figS7:tradeof}
\end{figure*}

By taking advantage of the quasi-resonant excitation scheme adopted in this work, we have been able to record $\gsl^{(2)}(\tau)$ traces with limited background emission at cryogenic temperature. Yet, as temperature increases, the thermal broadening of QD emission lines leads to an increased contribution of biexciton PL to the correlated counts and a reduction in the single-photon purity that follows accordingly. Thanks to the large biexciton binding energy of QDs emitting around \SI{4.5}{\ev}, single-photon emission could still be observed at RT, as described in the main text. We showed additionally that the resulting increase in $\gsl^{(2)}(0)$ can be accounted for by considering the biexciton luminescence as an uncorrelated background \cite{Brouri2000}:
\begin{equation}\label{eq3:uncorr}
\gsl^{(2)}(0) = 1 - \rho^2 \ \ ; \ \ \rho = \frac{I_X}{I_X + I_{XX}},
\end{equation}
where $I_X$ ($I_{XX}$) corresponds to the PL intensity of the exciton line (biexciton line) detected in the bandpass of the HBT interferometer. Both intensity terms can be determined by fitting the QD emission lines with separate peak functions (see inset of Fig.~\ref{figS7:tradeof}) and by integrating each component in the energy interval of the HBT setup. At \SI{300}{\kelvin} and for an excitation power density of \SI{400}{\kilo\wcm}, the single-photon purity of about $\gsl^{(2)}(0)=0.27$ estimated for QD$_{\rm A}$ using Eq.~\ref{eq3:uncorr} (see Fig.~\ref{figS7:tradeof}) is very close to the value $\gsl^{(2)}_{fit }(0)=\SI{0.3(1)}{}$ extracted from the second order autocorrelation function (see Fig.~4 in the main text). To strengthen this result, Eq.~\ref{eq3:uncorr} should ideally be tested for different bandpasses. Our setup is unfortunately limited to a bandpass of \SI{8}{\milli\ev}. Nevertheless, we can still infer the prospective evolution of the single-photon purity when including an increase in the spectral collection efficiency. Figure~\ref{figS7:tradeof} highlights the prediction we made based on Eq.~\ref{eq3:uncorr} and the exciton and biexciton intensities extracted by fitting the PL peaks. Thus, RT single-photon emission is foreseen to be maintained when collecting up to \SI{90}{\percent} of the exciton PL. In other words, beyond such an exciton PL collection efficiency, the spurious contribution of the biexciton line in the bandpass leads to $\gsl^{(2)}(0)\geq0.5$. Besides, we should point out the remarkable stability of $\gsl^{(2)}(0)$ up to a \SI{50}{\percent} collection efficiency of the exciton PL intensity, with $\gsl^{(2)}(0)=0.29$ for a bandpass of \SI{30}{\milli\ev}. Using such a bandpass would in turn enable us to perform measurements at lower excitation power density, reducing even more the contribution of the biexciton line. We thus expect our results to be amenable to improvement.

\section{Power-dependent Stark shift with polar GaN/AlN quantum dots}

\begin{figure*}
	\centering
	\includegraphics[width= 0.8\textwidth]{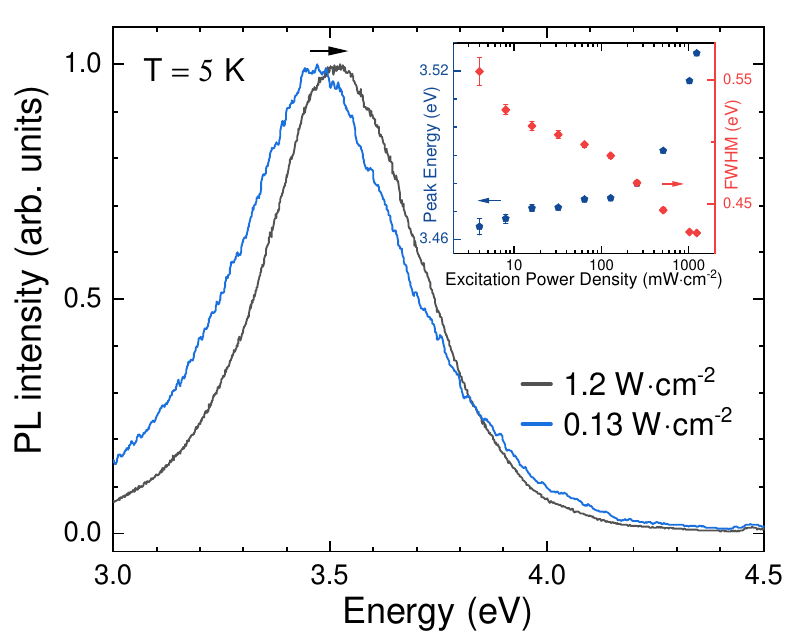}
	\caption{Low temperature ($T=$ 5 K) PL spectra of a polar GaN/AlN QD ensemble recorded at excitation power densities of 0.13 and \SI{1.2}{\watt\per\square\centi\meter}, respectively. The inset shows the blueshift of the peak energy and the FWHM narrowing of the PL spectra with increasing excitation power density.
	 }
	\label{figS8:Starkshift}
\end{figure*}

As mentioned in the main text, III-nitride QDs are especially sensitive to variations in the laser pump fluence when compared to their non-polar counterparts. Each electron-hole pair captured by a dot gets separated by the built-in electric field and contributes, in turn, to its screening. Hence, the higher the excitation power density, the larger the number of trapped excitons per dot and the lower the Stark shift induced by this built-in field. The absolute magnitude of this shift is strongly dependent on the QD height. Indeed, electron-hole pairs trapped in large QDs experience a larger separation than in their smaller counterparts. The signature of this phenomenon is clearly observed in the PL spectra measured on QD ensembles at different excitation power densities shown in Fig.~\ref{figS8:Starkshift}.  The discrepancy between the \SI{130}{\milli\watt\per\square\centi\meter} (blue line) and \SI{1.2}{\watt\per\square\centi\meter} (black line) QD ensemble PL curves is larger for energies below \SI{3.5}{\ev}, which is consistent with the increase in QD height. Similarly, it explains the observed narrowing in the PL peak FWHM with excitation power density (inset of Fig.~\ref{figS8:Starkshift}) as the energy separation between low and high energy QDs becomes smaller. This power-dependent effect has an impact on ensemble TRPL measurements as for a given detection bandpass the luminescence from multi-excitonic recombination events (short-lived component of the decays) and the late luminescence from single exciton recombination (long-lived component of the decays) will obviously originate from different QD sets.

\section{Evolution of the bi-exponential photoluminescence decay with temperature.}

In the main text, we reported on the bi-exponential dynamic that governs the exciton recombination whatever the QD emission energy. Several scenarios were postulated to account for this behavior. One plausible explanation relies on a PL signal arising from the two distinct QD exciton energy bright states. In this framework, the fast decay time $\tau_{\rm S}$ accounts simultaneously for the recombination of the high energy bright state B$_2$ and its relaxation toward  lower energy states, i.e., toward the low energy bright state B$_1$ and the exciton dark states (cf. Fig. 2(a) of the main text). Since the long decay time $\tau_{\rm L}$ remains constant with temperature (cf. Fig.~9 of the main text), it is hinting at the \textit{a priori} negligible impact of phonon-mediated processes for the states exhibiting this recombination dynamic. In other words, $\tau_{\rm L}$ is assumed to be mainly driven by the radiative recombination of B$_1$. To account for this temperature-independence of $\tau_{\rm L}$,
a quasi-equilibrium should take place between B$_1$ and the dark states that occurs on a much faster timescale than $\tau_{\rm L}$.

On the other hand, the weight of the fast component strongly depends on the refilling of B$_2$, which occurs through absorption of phonons with an energy matching the splitting between B$_2$ and lower energy states. Hence, the enhancement of the PL intensity originating from the high energy bright state can occur either through an increase in temperature or a reduction in the bright state energy splitting. The latter coincides with an increase in QD size and can be monitored through energy-dependent TRPL measurements.

\begin{figure*}
	\centering
	\includegraphics[width= 0.8\textwidth]{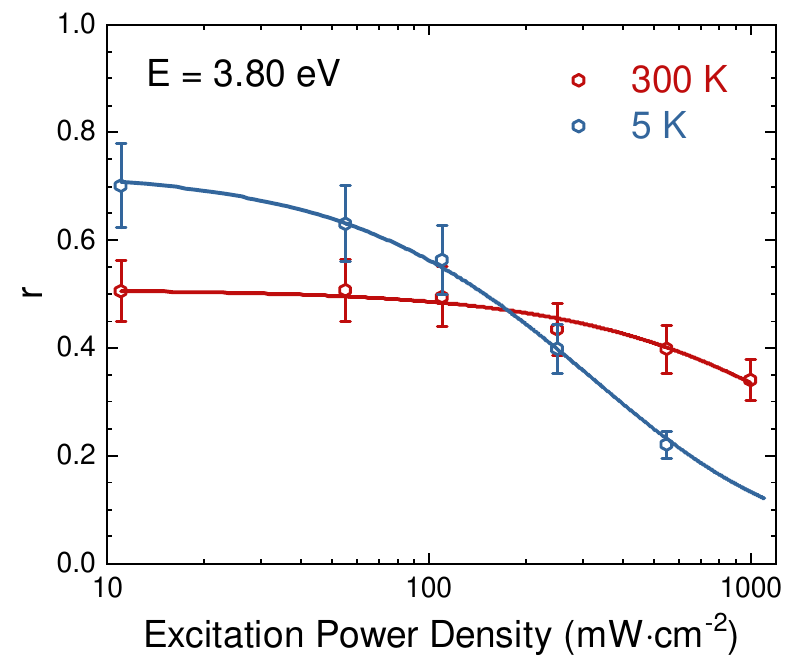}
	\caption{Evolution of the intensity ratio $r$ as a function of excitation power density for QDs emitting at \SI{3.80}{\ev}. Results collected at both 5 and \SI{300}{\kelvin} are given. The solid lines serve as a guide to the eye. The uncertainties for the $I_{\rm X}$ component are estimated by varying the lower bound when integrating the mono-exponential fit. Thus, the lower bound was changed by about the duration of the fast decay, starting from $\tau = 0$, and the extreme values of $I_{\rm X}$ are used to set the error bars. Each data point corresponds to the mean integrated $I_{\rm X}$ value.
	}
	\label{figS9:ratio3.8}
\end{figure*}

In order to test these predictions, we extracted the long-lived PL component of the TRPL transients by integrating the associated long-lived mono-exponential fit over the whole PL intensity decay profiles, (as shown in Fig.~8 of the main text). In each case, this component was normalized to the PL intensity decay profile integrated over the whole raw data (for delays $\tau > 0$), hence leading to the intensity ratio denoted by the letter $r$ in the main text. When reaching low excitation power densities, the contribution from multi-excitonic states to the TRPL transients is expected to vanish so that $r$ should saturate ($r(\tau\rightarrow 0) \rightarrow r_0$). This is shown in Fig.~\ref{figS9:ratio3.8} for QDs emitting at \SI{3.80}{\ev}. In the current framework, $r_0$ can be seen as a marker of the B$_1$ PL weight with respect to that of B$_2$, as shown by Eq.~(6) in the main text. In other words, the weight of B$_1$ should increase together with $r_0$.

\begin{figure*}
	\centering
	\includegraphics[width= 0.8\textwidth]{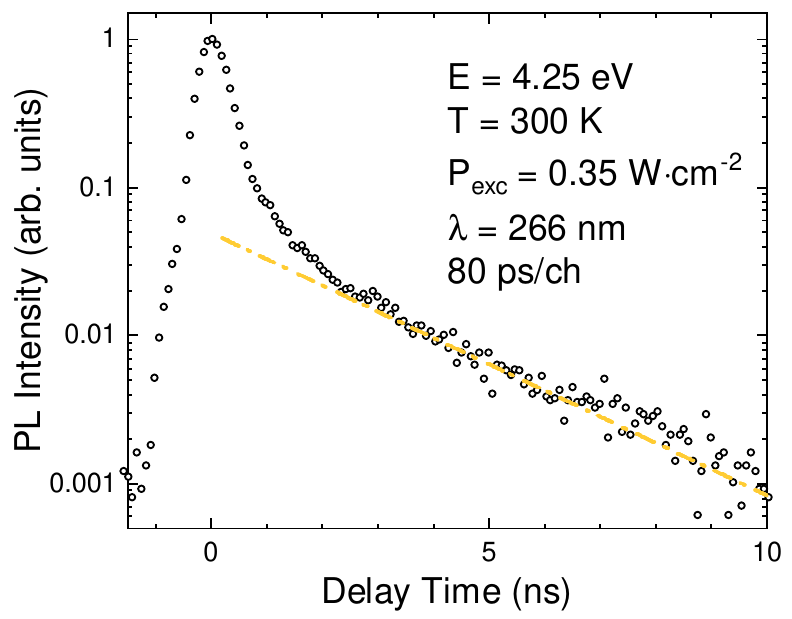}
	\caption{Photoluminescence intensity decay of GaN QDs emitting at \SI{4.25}{\ev} recorded at \SI{300}{\kelvin} for an excitation power density of \SI {0.35}{\wcm}. The tail is approximated by a single exponential with a decay time $\tau_{\rm decay} = \SI{2.45}{\ns}$ (yellow dash-dotted line).
	 }
	\label{figS10:transients4.25}
\end{figure*}

By monitoring a decrease in $r_0$ with decreasing QD emission energies (Fig.~10 of the main text), we showed that the long-lived decay contributes less to the PL of large QDs. This is in line with the expected decrease in the fine structure splitting with increasing QD size. From the ratios reported in Fig.~\ref{figS9:ratio3.8}, it appears now that $r_0$ is also decreasing with temperature for QDs emitting at \SI{3.8}{\ev}, as expected from the above-mentioned picture. The variation remains, however, relatively small (reduction of about \SI{30}{\percent}) and additional results are required to confirm the global trend. Similar measurements led on high energy QDs would prove extremely valuable. Indeed, the large fine structure splitting characteristic of small QDs results in a strong discrepancy between B$_1$ and B$_2$ recombination rates. At \SI{5}{\kelvin}, radiative recombination from the high energy bright state (X$_2$ line) is thus thermally suppressed and the X$_1$ prevails. On the other hand, X$_2$ dominates PL spectra as soon as B$_2$ refilling from lower energy states becomes thermally activated (cf. Fig.~\ref{figS4:Temperature}). From TRPL data, this should translate into a large ($r_0\rightarrow 1$) and a small ($r_0 \rightarrow 0$) $r_0$ value at \SI{5}{\kelvin}  and \SI{300}{\kelvin}, respectively. Given the very low luminescence signal issued from GaN/AlN QD ensembles for energies above \SI{4}{\ev}, we could not record any exploitable TRPL transients at RT under very low excitation power densities. Nevertheless, we could extract a ratio $r = 0.25$ from the RT TRPL transient collected at \SI{0.35}{\wcm} on QDs emitting around \SI{4.25}{\ev} (cf. Fig.~\ref{figS10:transients4.25}). The latter ratio represents a substantial decrease from the $r = 0.82$ value obtained at \SI{5}{\kelvin} under equivalent pumping conditions. Besides, $r$ is observed to saturate faster at high temperature, as shown for QDs emitting near 3.80 eV (Fig.~\ref{figS9:ratio3.8}). Hence, the ratio $r$ measured at \SI{300}{\kelvin} for any excitation power density should be closer to the $r_0$ limit than its \SI{5}{\kelvin} counterpart, i.e., $\frac{r}{r_0}(P_{\rm exc},\SI{300}{\kelvin}) > \frac{r}{r_0}(P_{\rm exc},\SI{5}{\kelvin}) \ \forall \ P_{\rm exc}$. At \SI{5}{\kelvin} and for QDs emitting at \SI{4.25}{\ev}, we determined $\frac{r}{r_0}(\SI{0.35}{\wcm}) > 0.9$. We can thus reasonably expect the ratio $r= 0.25$ measured at RT under identical excitation power density ($P_{\rm exc} = \SI{0.35}{\wcm}$) to differ by less than \SI{10}{\percent} from $r_0$. As such, $r_0(\SI{300}{\kelvin}) < 0.3$ for QDs emitting at \SI{4.25}{\ev}, which scales way below the $r_0(\SI{5}{\kelvin}) = 0.9$ determined for similar QDs at cryogenic temperature. This confirms that the short-lived TRPL component is thermally activated.

\section{Cathodoluminescence of the unetched GaN/AlN quantum dot sample}

\begin{figure*}
	\centering
	\includegraphics[width= 0.8\textwidth]{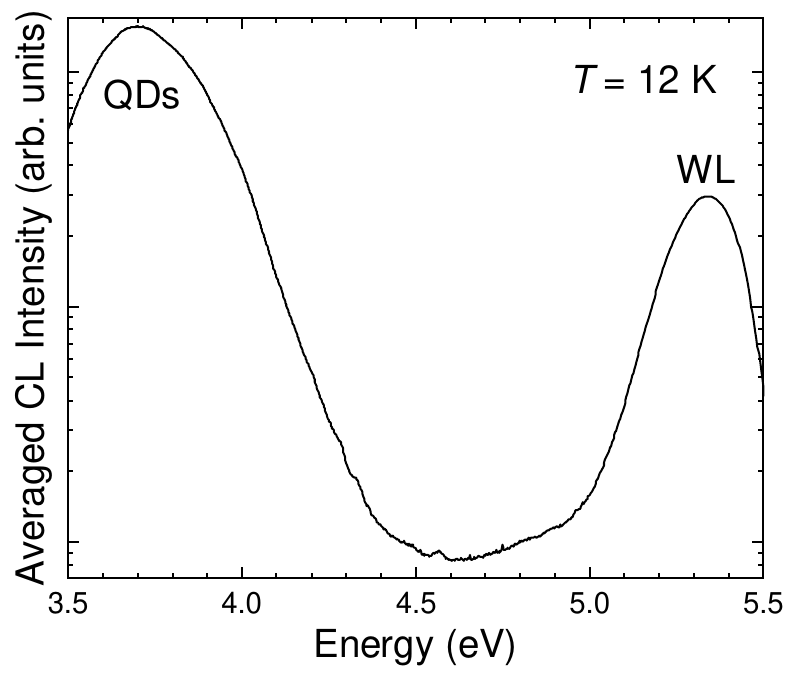}
	\caption{Low temperature averaged CL spectrum measured using an acceleration voltage of 6 kV over an unprocessed area of the GaN/AlN QD sample. The peak at \SI{5.3}{\ev} corresponds to an $\sim 1.5$ monolayer-thick GaN WL.
	 }
	\label{fig3:CL}
\end{figure*}

A representative cathodoluminescence (CL) emission spectrum recorded at $T=\SI{12}{\kelvin}$ using an acceleration voltage of 6 kV over an unetched area of the QD sample is shown in Fig.\ \ref{fig3:CL}. The WL emits around \SI{5.3}{\ev}, which corresponds to an equivalent GaN thickness of $\sim 1.5$ monolayers. Let us note that the maximum CL intensity of the QD ensemble is shifted by about \SI{200}{\milli\ev} with respect to the QD ensemble photoluminescence (cf. inset of Fig. 1\textbf{a} in the main text) as a result of the high carrier density generated in the present CL experiment, which leads to a screening of the built-in electric field. On the other hand, the WL emission remains marginally impacted even under high carrier injection because such very thin layer is almost insensitive to the quantum confined Stark effect.

\newpage

\bibliography{Light_Supp}